%% file: article.tex
\documentclass[preprint2]{aastex}

\usepackage{natbib}
\usepackage[utf8]{inputenc}
\usepackage[T1]{fontenc}
\usepackage{amsmath}
\usepackage{hyperref}

\shorttitle{Flows at the edge of an Active Region}
\shortauthors{Boutry et al.}

\begin{document}

\title{Flows at the Edge of an Active Region: Observation and Interpretation}

\author{C. Boutry\altaffilmark{1,2},
 E. Buchlin\altaffilmark{2,1},
 J.-C. Vial\altaffilmark{2,1},
  \and S. R\'egnier\altaffilmark{3}}
  \email{eric.buchlin@ias.u-psud.fr}

\altaffiltext{1}{Univ Paris Sud, Institut d'Astrophysique Spatiale,
  UMR8617, 91405 Orsay, France}

\altaffiltext{2}{CNRS, Institut d'Astrophysique Spatiale, UMR8617,
  91405 Orsay, France}

\altaffiltext{3}{Jeremiah Horrocks Institute, University of Central Lancashire, Preston, Lancashire, PR1 2HE, UK}

\begin{abstract}

  Upflows observed at the edges of active regions have been proposed as the source of the slow solar wind.
  In the particular case of Active Region (AR) 10942, where such an upflow has been already observed, we want to evaluate the part of this upflow that actually remains confined in the magnetic loops that connect AR\,10942 to AR\,10943.
  Both active regions were visible simultaneously on the solar disk and were observed by STEREO/SECCHI EUVI. Using Hinode/EIS spectra, we determine the Doppler shifts and densities in AR\,10943 and AR\,10942, in order to evaluate the mass flows. We also perform magnetic field extrapolations to assess the connectivity between AR\,10942 and AR\,10943.
  AR\,10943 displays a persistent downflow in \ion{Fe}{12}. Magnetic extrapolations including both ARs show that this downflow can be connected to the upflow in AR\,10942. We estimate that the mass flow received by AR\,10943 areas connected to AR\,10942 represents about 18\% of the mass flow from AR\,10942. We conclude that the upflows observed on the edge of active regions represent either large-scale loops with mass flowing along them (accounting for about one-fifth of the total mass flow in this example) or open magnetic field structures where the slow solar wind originates.
 
\end{abstract}


\input{introduction}

\input{ar10943}

\input{both_b}

\input{discussion}

\bibliographystyle{apj}
\bibliography{boutry}

\end{document}

%% file: introduction.tex
\section{Introduction}

The Sun interacts with the whole heliosphere, and in particular with the planets of the solar system, through the solar wind. As the plasma $\beta$ (ratio of the plasma pressure to the magnetic pressure) is low in the low corona, the dynamics of the plasma is dominated by the magnetic field
(frozen-in condition) implying that the plasma material is flowing along magnetic field lines. In particular, the fast solar wind follows open magnetic field lines from solar coronal holes to the interplanetary space but the sources of
the slow wind remain an open issue.

Fast and slow winds can be distinguished according to their speeds (around $600$ and $300\,\textrm{km}\cdot\textrm{s}^{-1}$ respectively) and their composition, but this is not the only difference between both these types. The fast wind is expelled from coronal holes, especially at the poles \citep{1973SoPh...29..505K}, and possibly from the intersections of chromospheric network boundaries in coronal holes \citep{1999Sci...283..810H}. 
The slow solar wind is not as well understood as the fast one, for various reasons probably
related to its time variability. Its transient nature and its relation with large-scale coronal structures,
have been revealed from both in-situ (e.g. \citealt{2009SoPh..256..327K}) and remote-sensing observations (e.g. the blobs
of \citealt{1997ApJ...484..472S}). As for the coronal sources, the edges of coronal holes (Coronal Hole Boundaries)
have been proposed as the location of reconnection between coronal hole (CH) and non-CH magnetic
fields, because of the differential rotation between these two kinds of regions \citep{2005ESASP.592..645S,2004ApJ...612.1196W}.
Recent spectrocopic and imaging observations with SUMER/SOHO and XRT/Hinode seem to support
this mechanism (\citealt{2004ApJ...603L..57M}, \citealt{2010A&A...516A..50S}).
However, other locations and mechanisms
have been proposed, such as streamer boundaries \citep{2010AdSpR..46.1400A} and the edges of Active Regions \citep{1999JGR...10416993K,2004SoPh..223..209L,ko06}. As we shall see below, this latter possibility has been very recently
put forward in the context of an Active Region close to an "open field" region, an issue we focus on
in this Paper.

Active regions (ARs) in the Sun's atmosphere are composed of closed multi-temperature loops in the solar corona.
Recently a specific flows distribution has been shown for some ARs. It corresponds to redshifts inside loops (\citealt{2008A&A...481L..49D}, \citealt{2009ApJ...694.1256T}) and blueshifts at the edge of active regions (\citealt{2007ApJ...667L.109D},  \citealt{2007Sci...318.1585S}, \citealt{2008ApJ...676L.147H}, \citealt{2009ApJ...705..926B},  \citealt{2011A&A...526A.137D}). In the last two papers, according to magnetic field extrapolations, blueshifts are observed along fanning out, far-reaching or even open field lines. Then the observed  flows seem to occur at the
boundaries between active regions and coronal holes.
Thus the outflows can supply mass to the solar wind, as suggested by scintillation measurements at 2.5 solar radii, these outflows coming from an actually open region at the edge of the active region \citep{1999JGR...10416993K}.

Active Region 10942 has already been extensively studied: Fast upflows have been observed at the North-East edge of this AR from Hinode/EIS Doppler shifts in \ion{Fe}{12} 195.12 \AA\ \citep{2008ApJ...676L.147H, 2009ApJ...705..926B, 2009ApJ...706L..80M} and from apparent flows in Hinode/XRT \citep{2007Sci...318.1585S} and TRACE \citep{2009ApJ...706L..80M} time series. These upflows have been proposed as a source of the slow solar wind. Linear force-free \citep{2008ApJ...676L.147H,2009ApJ...705..926B} and potential field source surface \citep{2007Sci...318.1585S} extrapolations show open field lines in agreement with this hypothesis.

\citet{2008ApJ...676L.147H} have also noticed that AR\,10942 was connected by large scale loops to a magnetic dipole, approximately 400\,arcsec away, which actually is AR\,10943. This is a clear evidence for magnetic connection, but matter exchange has never been quantified. Large scale loops connecting two active regions have been observed since Skylab \citep{1976RSPTA.281..365V} and more recently with SDO \citep{2011JGRA..11604108S}.
The connection between far-distance active regions has also been observed in the case of transequatorial loops whether they are elongated stable structures \citep{1999SoPh..187...33F,2001SoPh..202...81F} or ephemeral  loops related to flare and filament eruption \citep{2007SoPh..244...75W}. Of special interest here are the spectroscopic observations of a transequatorial loop which indicate that the loop plasma was multithermal and covered roughly 2 orders of magnitude in temperature \citep{2006ApJ...636L..57B}. Moreover line-of-sight steady flows of the order of 30 to 40\,km.s$^{-1}$ were detected and interpreted as a necessary condition for maintaining the loop structure.

The above discussions on flows in active regions should not lead us to forget the issue of the "rest" wavelengths used to define absolutely the flows, even if these flows are relatively important in AR. Because the region taken as a reference is often the quiet Sun (QS), the issue of average temperature-dependent  flows in the QS is critical. \cite{2011A&A...534A..90D} mention average line shifts at 1 MK < T < 1.8 MK bluer than those observed at 1 MK (about -1.8 $\pm$ 0.6 km.s$^{-1}$) translating into a maximum Doppler shift of -4.4 $\pm$ 2.2 km.s$^{-1}$  around 1.8 MK. If we assume that the actual uncertainties are of the order of 2 km.s$^{-1}$, one immediately sees that the sign itself of the velocities (flows) may be changed. This clearly shows the need of a very careful determination of the wavelength reference.

In this paper we set out to further explore the link between Active Regions 10942 and 10943. In Sec.~\ref{sec:ar10943} we use EUV image and spectroscopic observations to analyze the flows in AR\,10943. A special attention has been paid to the determination of the velocities taking account of the global flow velocities dependance on temperature in the solar atmosphere \citep{1998ApJS..114..151C, 1999ApJ...522.1148P}. To understand the magnetic connection between both regions, we compute magnetic field extrapolations in Sec.~\ref{sec:both}. Finally, the results are discussed in Sec.~\ref{sec:discussion} and we conclude in Sec.~\ref{sec:conclusion}.

%% file: ar10943.tex
\section{Observations of AR\,10943 and AR\,10942}
\label{sec:ar10943}

\subsection{Observation with STEREO~B/SECCHI EUVI image}
\label{sec:imobs}

\begin{table*}[htbp]
  \caption{\label{tab:list} List of observations on 2007 February 20: Field-of-view (FOV), heliocentric center of FOV, start times ($T_{\textrm{start}}$) or reference times ($T_{\textrm{ref}}$), end times ($T_{\textrm{end}}$), and exposure times ($T_{\textrm{exp}}$), center wavelength ($\lambda$). Start and end times correspond to the period during which raster observations are used.}
  \centering
  \begin{tabular}{lllllll}
    \hline \hline
    Instrument & FOV (arcsec) & Center (arcsec) & $T_{\textrm{start}}$ or $T_{\textrm{ref}}$ & $T_{\textrm{end}}$ & $T_{\textrm{exp}}$ & $\lambda$ or filter \\ \hline
    STEREO/SECCHI EUVI & Full disk   &  & 06:35 UT & 06:35 UT & 2\,s & 195\AA \\
    SoHO/MDI & Full disk   &  & 08:05 UT &  & 300s & 6767.8\AA \\
    Hinode/EIS & $128\times512$  & (93, -37)  & 05:47 UT  & 07:59 UT  & 60\,s & 195\AA  \\
    Hinode/EIS & $41\times400$  & (-464, -100)  & 11:16 UT  & 11:37 UT  & 30\,s & 195\AA  \\ \hline
  \end{tabular}
\end{table*}

\begin{figure*}[htbp]
\centering
 \includegraphics[width=1\linewidth]{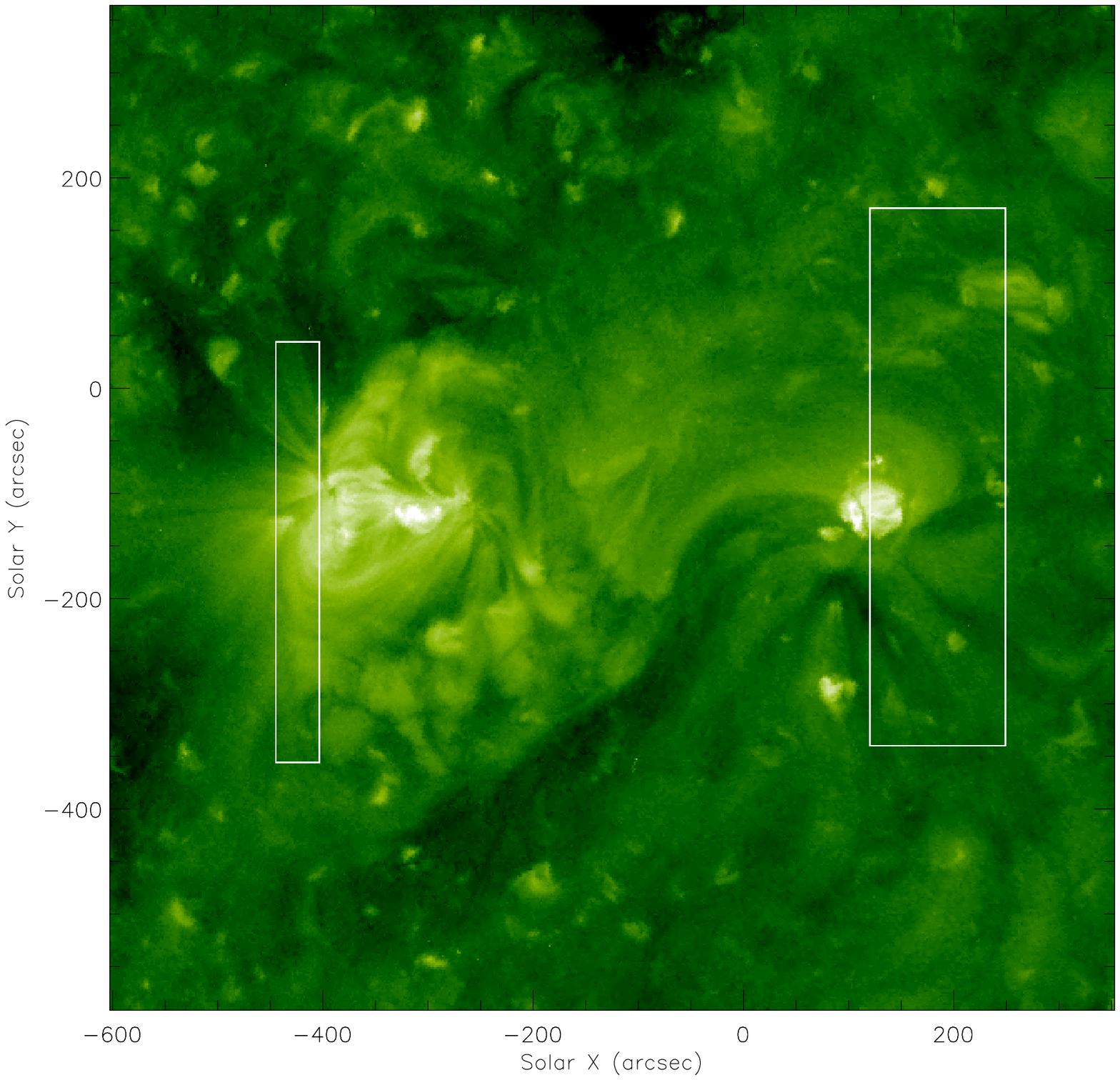}
 \includegraphics[width=1\linewidth]{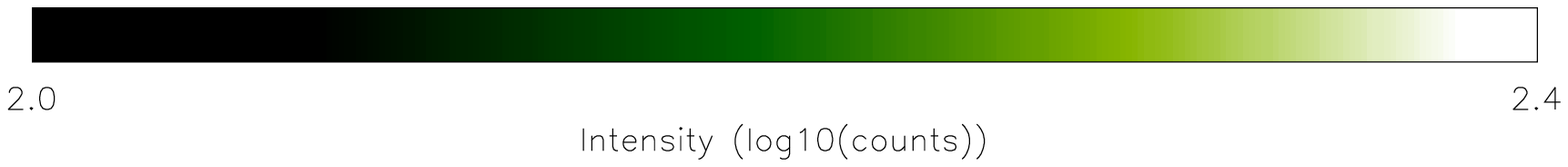}
  \caption{\label{fig:stereo} STEREO~B/SECCHI EUVI observations of AR\,10942 (on the left) and AR\,10943 (on the right) taken at 195\,\AA\ on 2007 February 20 at 8:05. The Hinode/EIS FOV at 05:47–07:59 UT for AR\,10943 and at 11:16-11:37 UT for AR\,10942 for the data analyzed in this paper are drawn as rectangles.}
\end{figure*}

\begin{figure*}[htbp]
\includegraphics[width=1\linewidth]{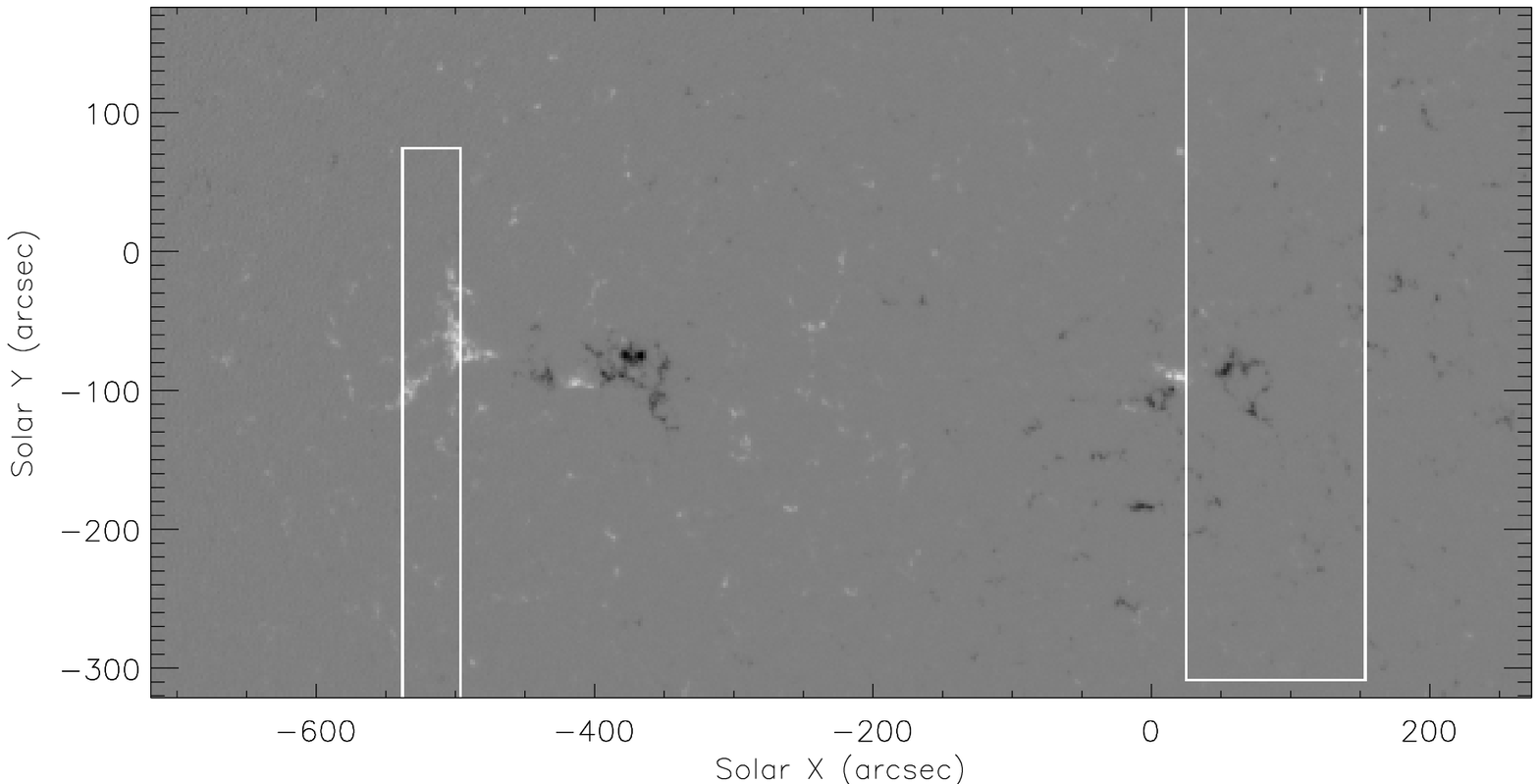}
\includegraphics[width=1\linewidth]{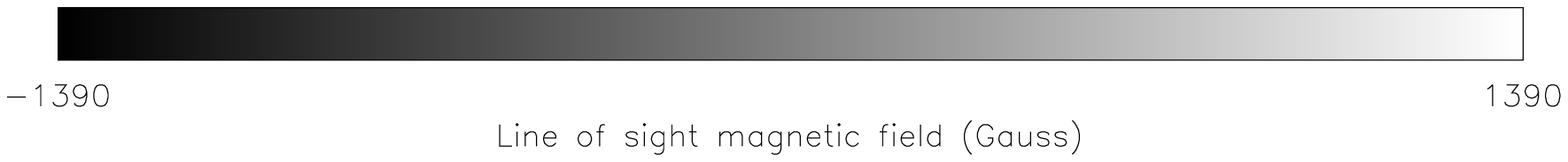}
  \caption{\label{fig:MDI_full}Full-disk SoHO/MDI magnetic field measurements on 2007 February 20 at $T_{ref}=8:05$~UT clipped to the area used for the magnetic field extrapolations. Both AR\,10942 (left) and 10943 (right) are visible. The Hinode/EIS FOV (at 05:47–07:59~UT for AR\,10943 and at 11:16-11:37~UT for AR\,10942, see Sec.~\ref{sec:secobs}) are drawn as rectangles.}
\end{figure*}

On 2007 February 20, less than four months after its launch, the heliocentric separation angle between the STEREO~B probe and the Earth was still negligible ($0.1\,\deg$). This means that STEREO~B/SECCHI EUVI images \citep{2008SSRv..136....5K,2008SSRv..136...67H} have the same viewing angle than SoHO and Hinode and that we can use these images in combination with SoHO/MDI and Hinode/EIS. STEREO/SECCHI EUVI was in its normal mode, with full-disk observations at a cadence of 10\,min in the EUV channel $\lambda$195\AA. We selected the 8:05~UT observation corrected by \verb+EUVI_prep+ 
from the SolarSoft library,
 shown in Fig.\ref{fig:stereo}.

In this image, both active regions can be seen simultaneously. 
On the eastern side, the EUVI image in 195\,\AA\ displays the AR\,10942 complex made of a set of loops connecting the two extreme polarities (see Fig.~\ref{fig:MDI_full}) of the AR, mainly in the southern side. Some straight (mostly fan-like) structures are also clearly seen on the north-eastern side (X\,=\,-400, Y\,=\,-50 to 0, see Fig.~\ref{fig:stereo}) which are candidate for open magnetic fields. At South, below AR\,10942, internal loops, rather compact structures (X\,=\,-300, Y\,=\,-300) seem to be the feet of (apparently) very sheared loops. On the western side of the image, another smaller and compact AR (10943) does not seem to be connected with its neighbouring regions, except for a set of diffuse loops at the East of AR\,10943 whose feet seem to be located in between the two ARs. One also notes that the lower side of these diffuse loops is very sharp. The overall picture is that the two ARs seem to be very disconnected on one hand, and that the eastern feet of the diffuse loops mentioned above could coincide with the QSL labelled “e” in Fig.4 of \cite{2009ApJ...705..926B}, on the other hand. The EUVI image in the hot 284\AA\ line confirms this picture, contrary to the \ion{He}{2} image where the chromosphere between the two AR does not seem to be very perturbed. Finally, the 171 image is more puzzling because it does not display the above (too) hot loops but also shows some dark area on the western side of AR\,10942, which could be a coronal hole or a filament channel.

\subsection{Spectroscopic observations}
\label{sec:secobs}

\begin{figure}[htbp]
  \centering
  \hspace*{3mm}\includegraphics[width=0.44\linewidth]{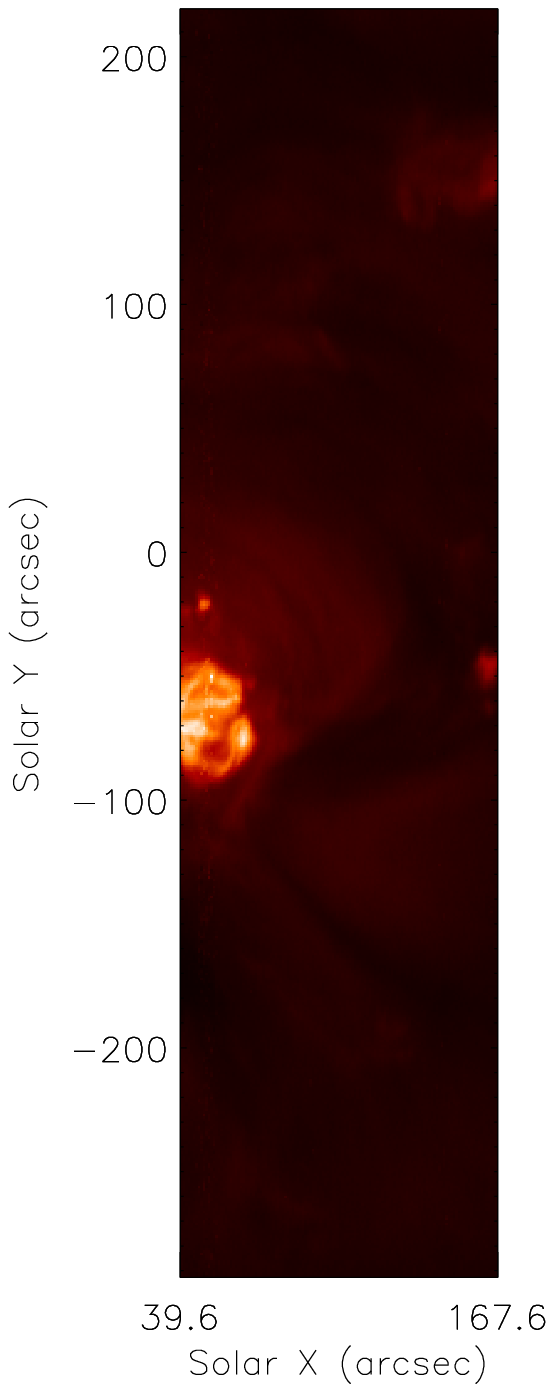}
  \hspace*{3mm}\includegraphics[width=0.44\linewidth]{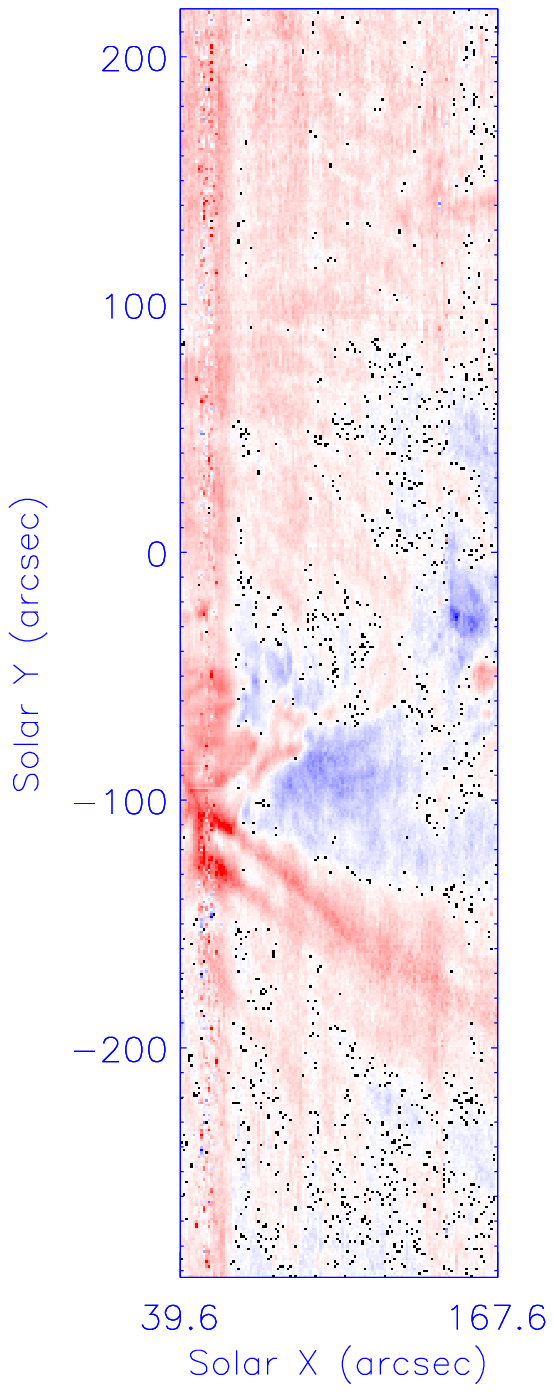}

  a)\includegraphics[width=0.44\linewidth]{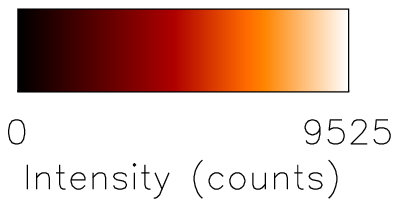}
  b)\includegraphics[width=0.44\linewidth]{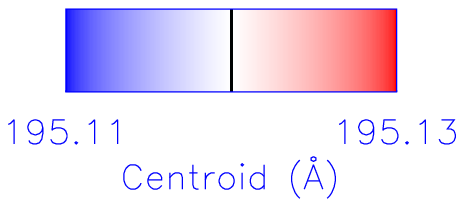}

  \centering
  \hspace*{3mm}\includegraphics[width=0.44\linewidth]{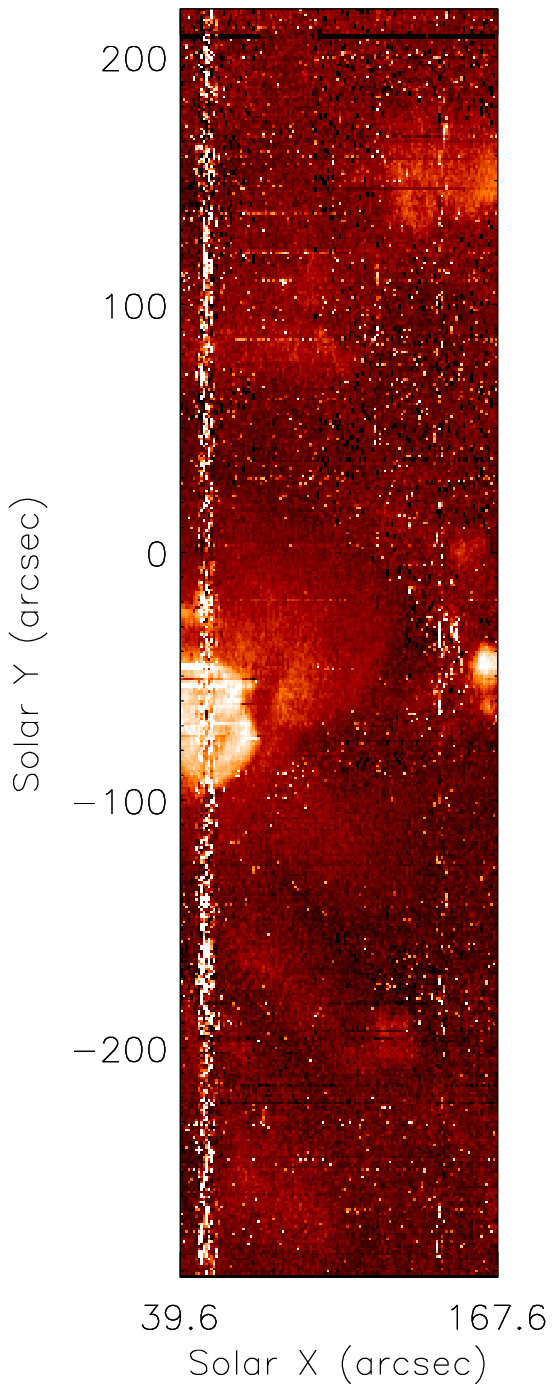}
  \hspace*{3mm}\includegraphics[width=0.44\linewidth]{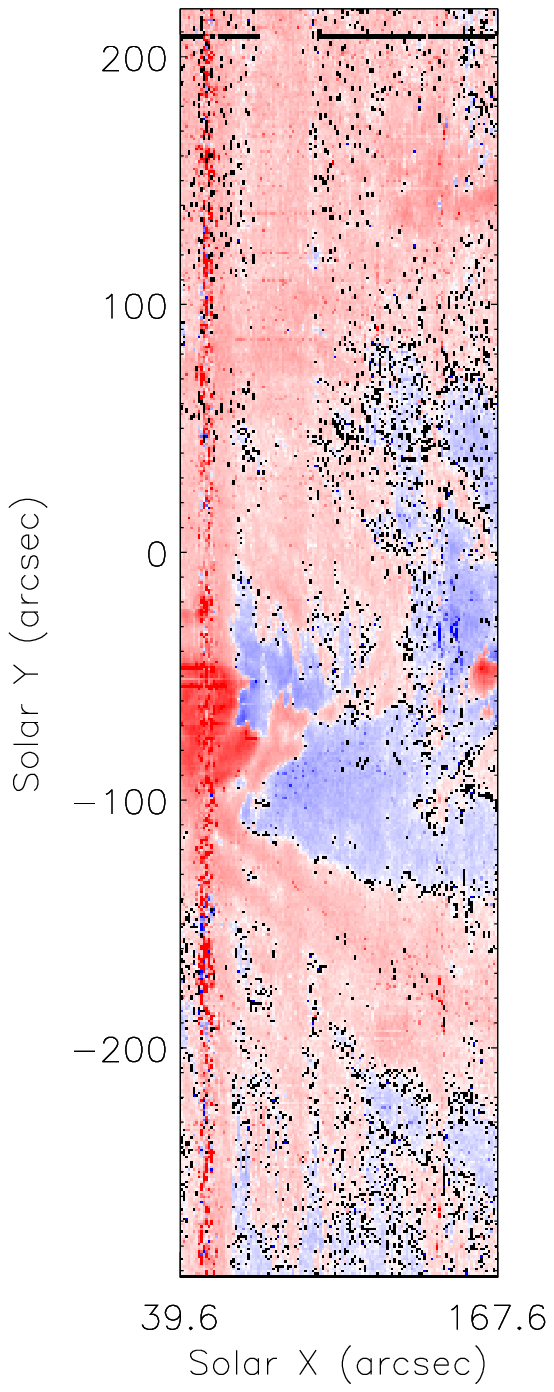}

  c)\includegraphics[width=0.44\linewidth]{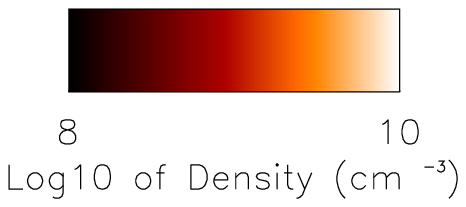}
  d)\includegraphics[width=0.44\linewidth]{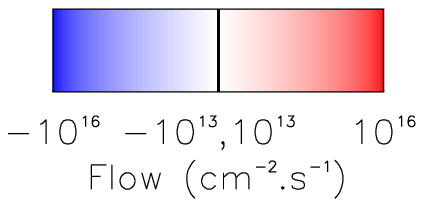}
  
  \caption{\label{fig:eis43}
    Maps of AR\,10943 observed on 2007 February 20 at 05:47–07:59 UT with Hinode/EIS in \ion{Fe}{12} 195.12\,\AA:  a) Intensity (integrated on the wavelength dimension); b) Centroid wavelength; c) density from the \ion{Fe}{12} 196.64\,\AA / \ion{Fe}{12} 195.12\,\AA\ line ratio; d) Mass flow  through a surface perpendicular to the line of sight.
  }
\end{figure}

\begin{figure}[htbp]
  \centering
  \hspace*{3mm}\includegraphics[width=0.44\linewidth]{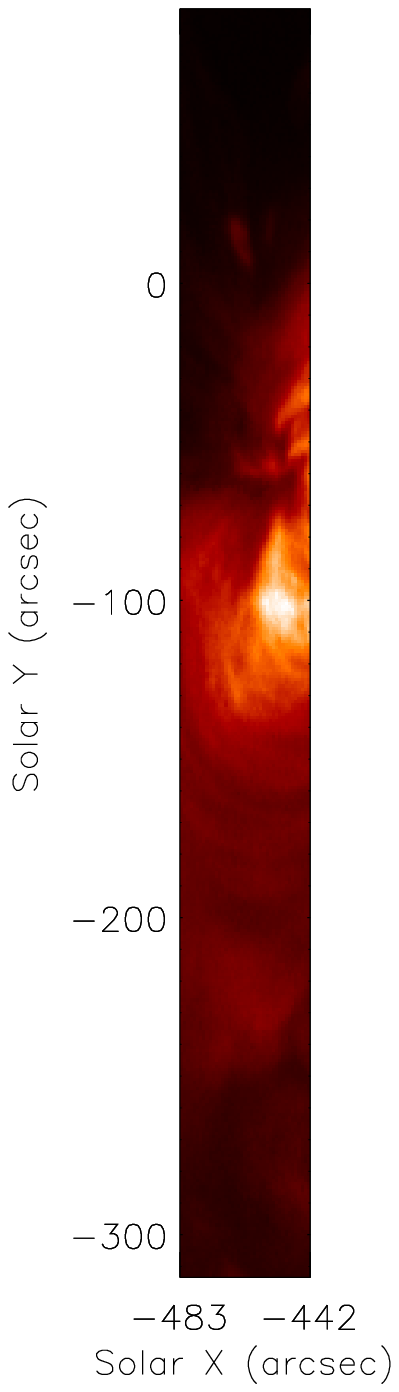}
  \hspace*{3mm}\includegraphics[width=0.44\linewidth]{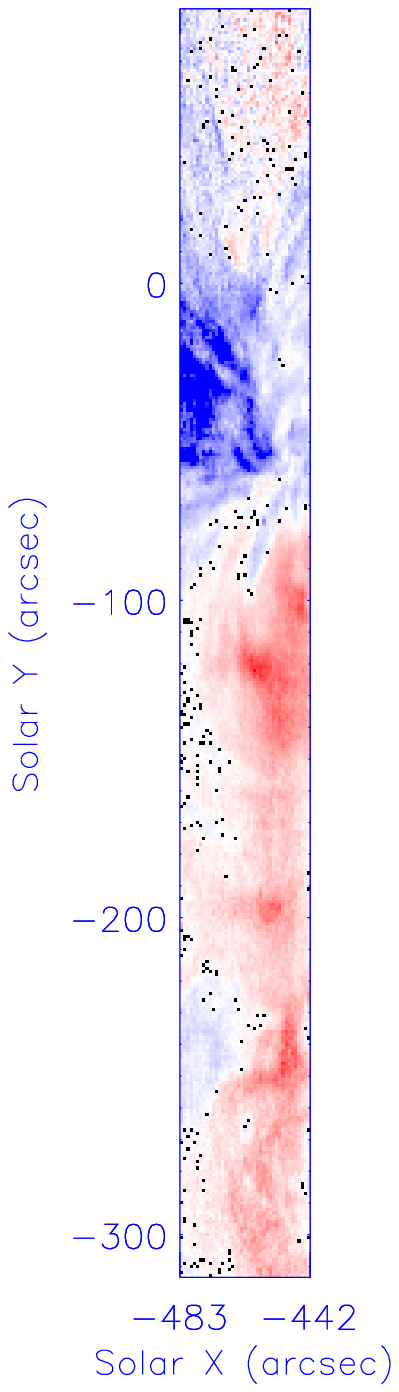}

  a)\includegraphics[width=0.44\linewidth]{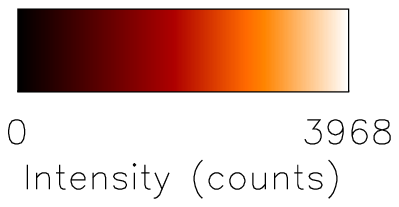}
  b)\includegraphics[width=0.44\linewidth]{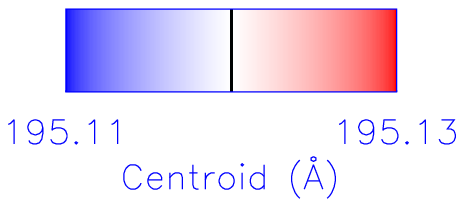}

  \centering
  \hspace*{3mm}\includegraphics[width=0.44\linewidth]{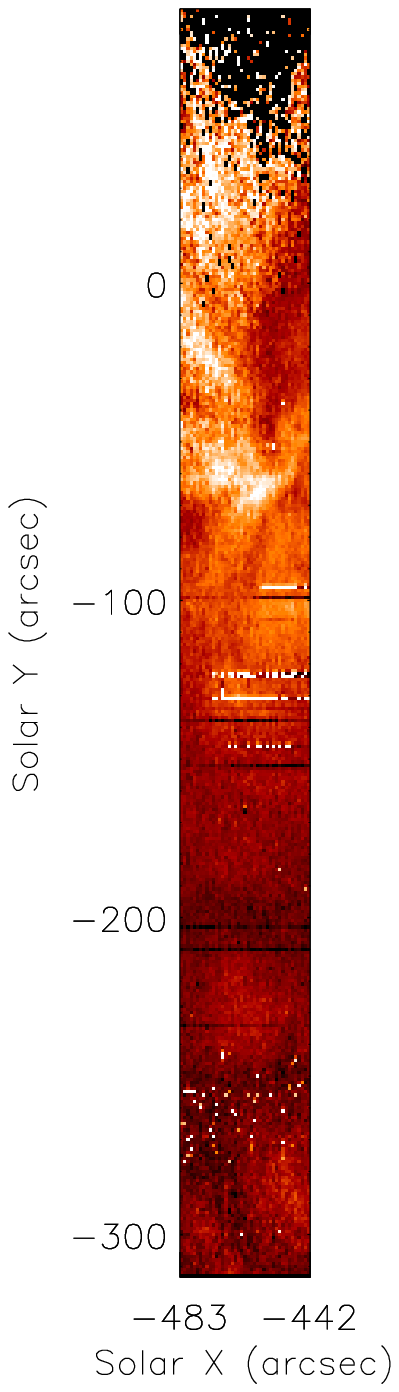}
  \hspace*{3mm}\includegraphics[width=0.44\linewidth]{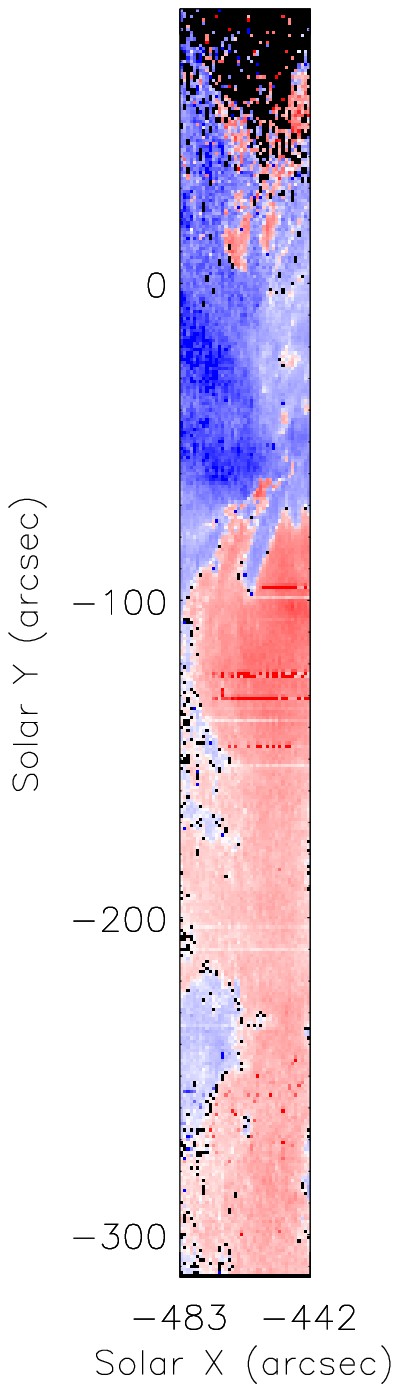}

  c)\includegraphics[width=0.44\linewidth]{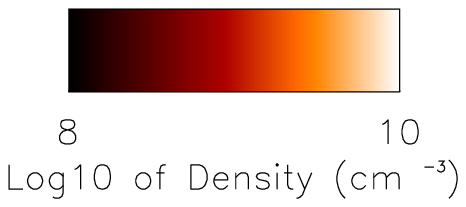}
  d)\includegraphics[width=0.44\linewidth]{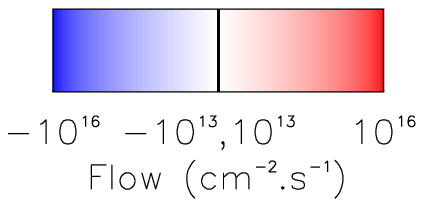}
  
  \caption{\label{fig:eis42}
    Maps of AR\,10942 observed on 2007 February 20 at 11:16-11:37 UT with Hinode/EIS in \ion{Fe}{12} 195.12\,\AA:  a) Intensity (integrated on the wavelength dimension); b) Centroid wavelength; c) density from the \ion{Fe}{12} 196.64\,\AA / \ion{Fe}{12} 195.12\,\AA\ line ratio; d) Mass flow  through a surface perpendicular to the line of sight.
  }
\end{figure}

\begin{figure}[htbp]
\centering
 \includegraphics[width=1\linewidth]{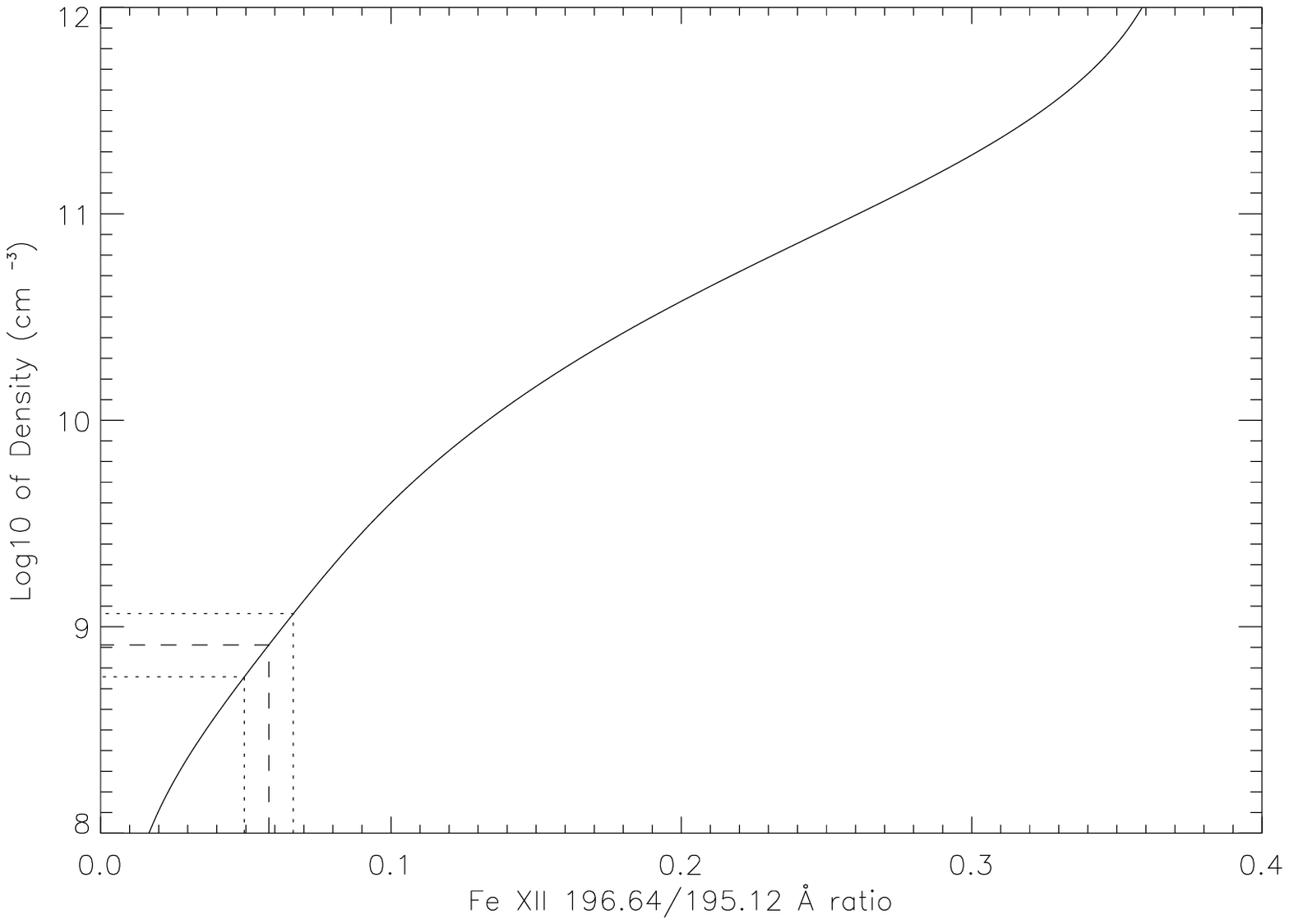}
  \caption{\label{fig:density_ratio} Density as a function of the \ion{Fe}{12} 196.64\,\AA /\ion{Fe}{12} 195.12\,\AA\ intensity ratio according to CHIANTI \citep{2009A&A...498..915D}. The dashed lines represent the average of the intensity ratio and the corresponding density. The dotted lines represent the error bars on the intensity ratio computed from the uncertainties on the measurements of intensities and the related density error bars. }
\end{figure}

Spectroscopic information can be obtained from the Hinode/EUV imaging spectrometer (EIS; \citealt{2007SoPh..243...19C}). We select two raster scans on 2007 February 20 that covers part of AR\,10943 (at 05:47\,UT, study ID 45) and part of AR\,10942 (at 11:16\,UT, study ID 57) respectively in order to have full spectra (around selected spectral lines) as a function of both solar dimensions in this region. The slit positions during these raster scans are shown in the STEREO image (Fig.~\ref{fig:stereo}) and in Table~\ref{tab:list}.
Both raster scans are partial on the active regions. Nevertheless, as we show with magnetic field extrapolation in Sec~\ref{sec:both}, the feet of the interconnecting loops between the two active regions are located at the respective edges of the active regions in the FOV of EIS. So the raster scans are sufficient for our study.
The delay of the scans and between the scans are negligible in comparison with the time of the continuous flows which are visible on a few days.

We apply the correction procedures \verb+eis_prep+ and \verb+eis_slit_tilt+ from the SolarSoft library. An additional correction must be applied for the orbital temperature variation of EIS; for this we have chosen to develop a specific method which is described in Appendix~\ref{sec:app} : we use the orbital variation in the \ion{He}{2} 256.32\,\AA\ line to correct the orbital lineshifts in other lines. Indeed, this line is chromospheric and optically thick, the insensitivity of its centroid with respect to activity allows us to better isolate orbital variations. 

We focus on the \ion{Fe}{12} 195.12\,\AA\ line which is emitted around $\log T = 6.1$. We produce intensity (Fig.~\ref{fig:eis43}a for AR\,10943 and Fig.~\ref{fig:eis42}a for AR\,10942) and Doppler velocity (Fig.~\ref{fig:eis43}b for AR\,10943 and Fig.~\ref{fig:eis42}b for AR\,10942) maps deduced from the parameters of a single Gaussian fit of this line using the correction of orbital variation. We choose to ignore the self-blending of \ion{Fe}{12} 195.12\,\AA\  with \ion{Fe}{12} 195.18\,\AA\ because we do not focus on the width (which is the most influenced parameter) and our study concerns low density structures where the contribution of \ion{Fe}{12} 195.18\,\AA\ is negligible \citep{2009A&A...495..587Y}.

The level 2 data show that the velocities in the active region are of the same order in the other EIS windows available except for \ion{Fe}{13} 196.54\,\AA\ where that pattern is reversed. In the \ion{Fe}{12} 186.88\,\AA\ and  \ion{Ca}{17} 192.82\,\AA\ lines, the core of the AR is blue but the other structures are the same. We cannot conclude about the \ion{Fe}{13} 196.54\,\AA\ and \ion{Fe}{12} 186.88\,\AA\ lines velocities because they are far different from the \ion{Fe}{13} 202.04\,\AA\ and \ion{Fe}{12} 195.12\,\AA\ respectively. The automatic analysis could not be sufficient. The \ion{Ca}{17} 192.82\,\AA\ line shows that the flow is upward in the core of the AR at very high temperature ($\log T_{max}=6.7$) but there are still downflows in the vicinity of the active region.

However the flow rate through a surface perpendicular to the line-of-sight is higher there than in the rest of the whirl. These downflows are persistent for a few days, therefore an impulsive event such as a jet is excluded as a viable mechanism producing these localised redshifts.

One also notes a redshifted area in the top third of the raster FOV. This area is not studied here for two reasons: it is not magnetically connected to AR\,10942 (see Sec.~\ref{sec:extra}) and the redshift is rather weak (a few km.s$^{-1}$).

We get a density map (Fig.~\ref{fig:eis43}c for AR\,10943 and Fig.~\ref{fig:eis42}c for AR\,10942) by also fitting the \ion{Fe}{12} 196.64\,\AA\ line, and computing the \ion{Fe}{12} 196.64\,\AA  / \ion{Fe}{12} 195.12\,\AA\ intensity ratio, which is sensitive to density. The density is deduced from the theoretical intensity ratio produced by the CHIANTI atomic database \citep{1997A&AS..125..149D,2009A&A...498..915D}, as shown in Fig.~\ref{fig:density_ratio}.

In order to derive the flow rate through a surface perpendicular to the line-of-sight (Fig.~\ref{fig:eis43}d for AR\,10943 and Fig.~\ref{fig:eis42}d for AR\,10942), we multiply the density by the Doppler velocity.

The core of AR\,10943 is characterized by hot loops bright in intensity (with a maximum of 9500 counts/pix, Fig.~\ref{fig:eis43}a), downward Doppler velocities and high densities between $4\times10^9$\,cm$^{-3}$ and $1\times10^{10}$\,cm$^{-3}$. These densities are consistent with values in active region loops by \cite{2008ApJ...686L.131W} ($1.3\times10^9$ and $9.5\times10^{10}$\,cm$^{-3}$ for \ion{Fe}{12}) and with \cite{2009A&A...495..587Y} ($3\times10^8$\,cm$^{-3}$ $\le n_e \le$ $1\times10^{11}$\,cm$^{-3}$).
The whirl of faint plasma around the core of AR\,10943 is mostly blueshifted, with electron density between $6.3\times10^8$ and $1\times10^9$\,cm$^{-3}$; moreover there is a clear straight redshifted (up to 16 km.s$^{-1}$) structure cutting the whirl in the South-East edge where the density is notably low (around $5\times10^8$\,cm$^{-3}$) but higher than Quiet Sun densities for \ion{Fe}{12} ($n_e=2.5-3.2\times10^8$\,cm$^{-3}$,  \citealt{2009ApJ...700..762W}).

%% file: both_b.tex
\section{Magnetic connection between AR\,10943 and AR\,10942}
\label{sec:both}

\subsection{Magnetic field observations}
\label{sec:magobs}

We investigate the possible magnetic connectivity between the two active regions
AR\,10942 and AR\,10943. We use a SOHO/MDI level 1.8 96 minutes line-of-sight magnetogram (see Fig.\ref{fig:MDI_full}) to study the
distribution of the photospheric magnetic field. The SOHO/MDI magnetograms have
been recorded on 2007 February 20 at $T_{ref}=08:05$\,UT (between the raster scans analysed in Sec.~\ref{sec:ar10943} and simultaneous with the STEREO SECCHI/EUVI image of fig.~\ref{fig:stereo}). We select an area encompassing the active regions (heliocentric coordinate $x$ from -720 to 275\,arcsec and $y$ from -321 to 177\,arcsec). The total unsigned magnetic
flux for this area is $4.05\times 10^{22}$ Mx and the flux unbalance is only about 1.7\% (a very low value for extrapolation). The
total unsigned flux for AR\,10942 is $9.90\times 10^{21}$ Mx with a negative flux of
$4.44\times 10^{21}$ Mx and a positive flux of $5.46\times 10^{21}$ Mx. For AR\,10943, the
total unsigned flux is $5.53\times 10^{21}$ Mx with a negative flux of $4.12\times 10^{21}$
Mx and a positive flux of $1.41\times 10^{21}$ Mx. The net magnetic flux for AR\,10943
is about 50\% of the total flux. AR\,10943 is in excess of negative flux, while
AR\,10942 is in excess of positive flux. If a magnetic connection exists between
the two active regions, thus this connection is between the positive polarity of
AR\,10942 and the negative polarity of AR\,10943.

\subsection{Magnetic field extrapolation}
\label{sec:extra}
\newlength{\figbwidth}
\setlength{\figbwidth}{\linewidth} 
\begin{figure}[htbp]
  \rlap{\smash{\raisebox{-2em}{(a)}}}\includegraphics[width=\figbwidth]{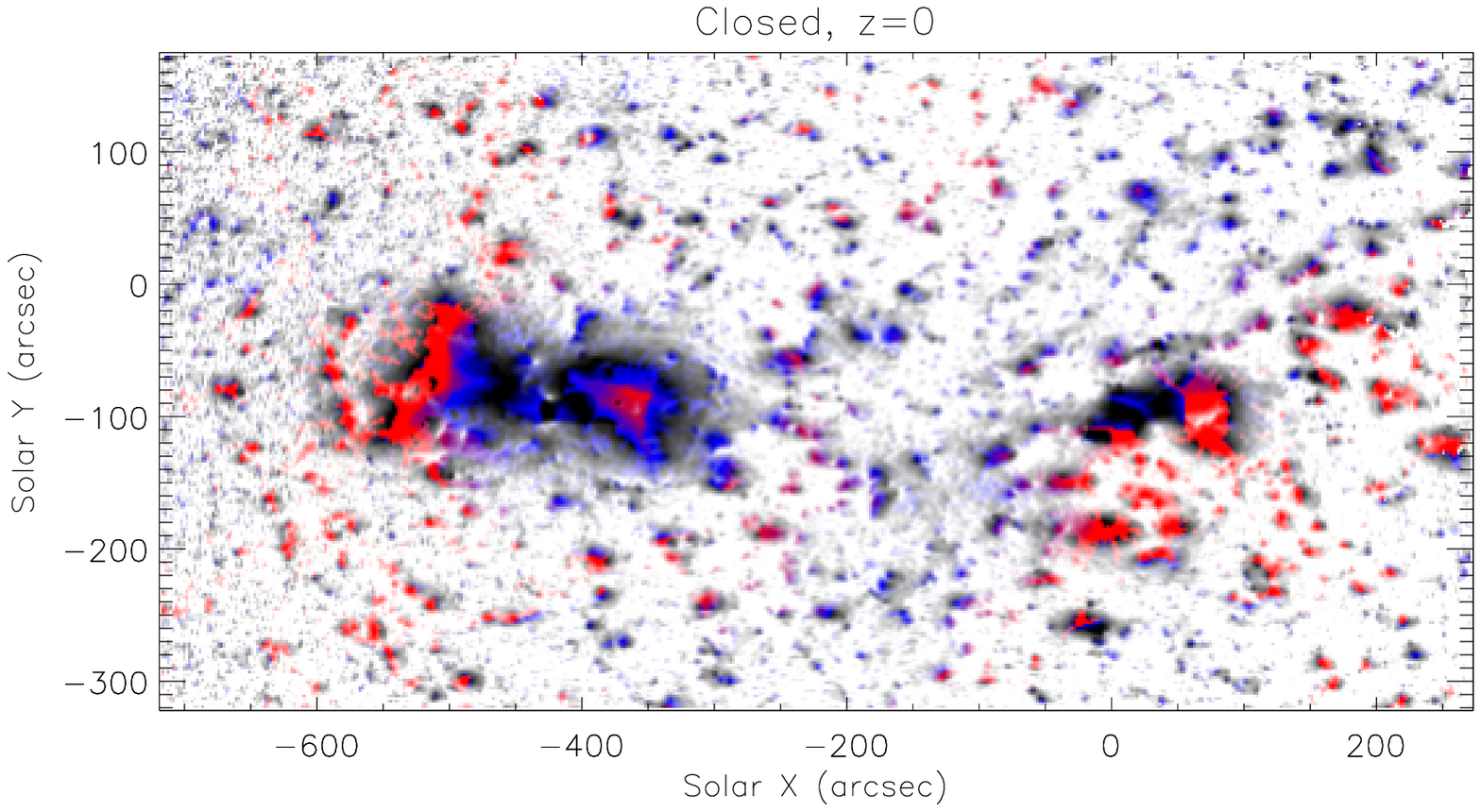}
  
  \rlap{\smash{\raisebox{-2em}{(b)}}}\includegraphics[width=\figbwidth]{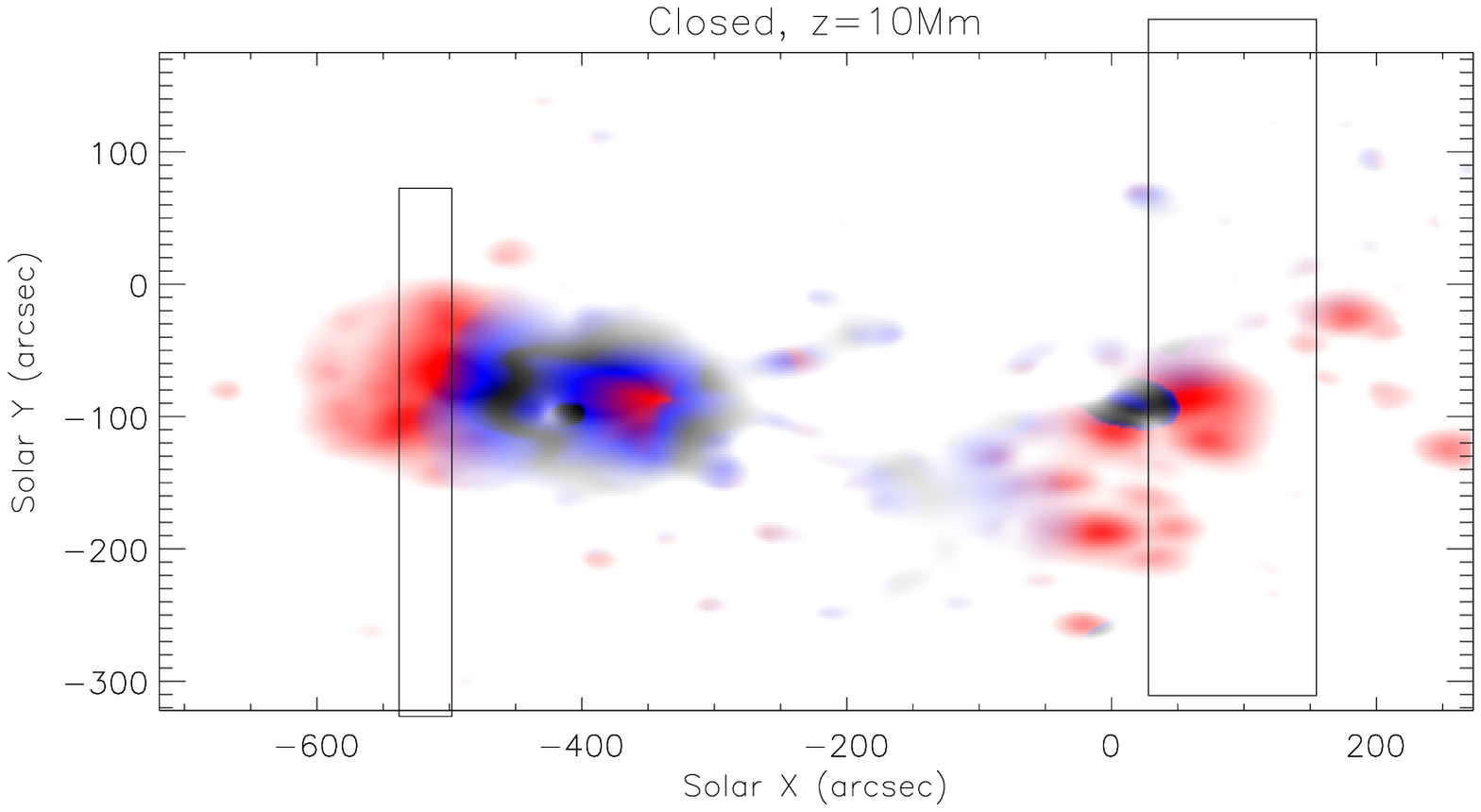}
  
  \rlap{\smash{\raisebox{-2em}{(c)}}}\hfill\includegraphics[width=.6\figbwidth]{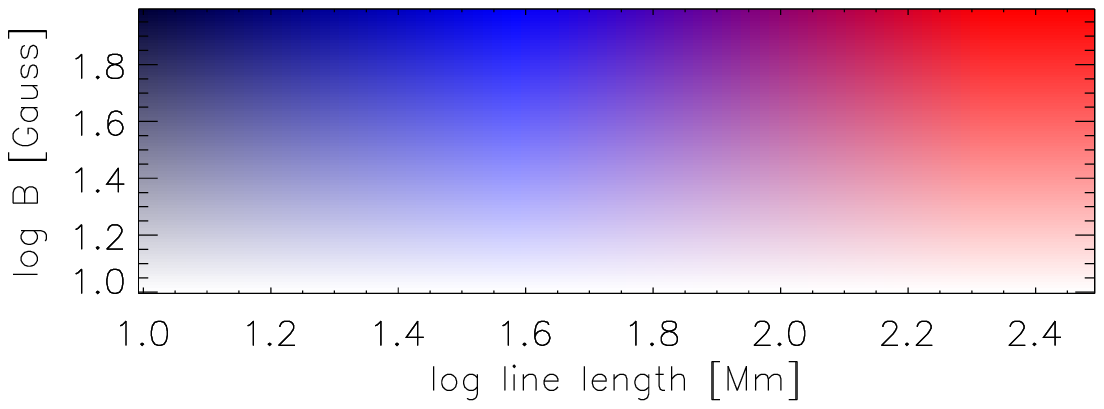}\hfill\llap{\hphantom{(c)}}

  \caption{\label{fig:connect} Connectivity maps for the potential field model in the case of closed boundary conditions, at two different altitudes: at the photosphere $z = 0$ (a) and at $z = 10$Mm (b). The color in any given pixel represents the length of the closed magnetic field line starting from this point (at the given altitude) and the modulus of the magnetic field at this position, according to the caption in panel (c). The magnetic field is color-coded from 10 to 100\,G and saturated below and above these values, while the the length is coded from 10 to 300\,Mm and saturated below and above these values. Rectangles in (b) indicate the EIS FOV.}
\end{figure}

To study the connectivity between the two active regions, we derive the 3D
magnetic field based on a potential field extrapolation
\citep[e.g.,][]{sch64,sem68}. The potential field assumption cannot describe the
effects of electric current on the magnetic field lines such as shear arcades or
twisted flux bundles. However, the potential field assumption is a good estimation of the
large-scale connectivity of the magnetic field 
\citep[see.g.,][]{sch69,lev82,wan92}. In addition, \citet{reg11a} showed that,
by comparing several magnetic field assumptions (potential, linear force-free,
and nonlinear force-free fields), the magnetic topology and connectivity are
well preserved when the current density is modified, which justifies the
use of the potential model in this study. 

The potential field extrapolation method solves the following equation:
\begin{equation}
\vec \nabla \times \vec B = \vec 0,
\end{equation}
with the bottom (or photospheric) boundary condition given by the vertical
component of the magnetic field provided by the SOHO/MDI magnetogram selected area (see Sec.~\ref{sec:magobs}). The
numerical technique relies on the computation of the scalar potential associated
to the magnetic field satisfying the Laplace equation. We perform the
potential field extrapolation for closed
boundary conditions for which the normal component of the magnetic field
vanishes on the sides and top boundaries. For the use of the SOHO/MDI
magnetogram as boundary conditions, we need to put constraints:
\begin{itemize}
\item[(i)] the center of the FOV is near the disk center and the
active regions are in a range of longitudes between E30 and W10, therefore the
line-of-sight magnetic field is not subject to projection effects;
\item[(ii)] the line-of-sight magnetic field component $B_s$ is converted into the
vertical (radial) component $B_z= B_s / \cos\theta$
where $\theta$ is the angle between the line-of-sight and the normal to the surface.
\item[(iii)]{as the magnetic field component satisfies the two items above,
we assume that the curvature of the Sun does not affect the connectivity of the
field. We thus compute the potential field in Cartesian coordinates.}
\item[(iv)]{as the magnetic flux through the surface containing the two active
regions is nearly balanced ($<$2\%), we do not force the magnetic flux to be
strictly balanced.}
\end{itemize}

In Fig.~\ref{fig:connect}, we plot the connectivity maps for the potential
model with closed boundary conditions: each pixel is color-coded as a function of the length of the field line associated with the photospheric footpoint and as a function of the magnitude of the field, according to the caption in panel (c). The connectivity maps have been computed at two different altitudes: at the photosphere $z = 0$ (a) and at $z = 10$\,Mm (b). The modulus of the magnetic field is below 10\,G in the Quiet Sun, mostly between 10\,G and 100\,G in the active regions at 10\,Mm while superior to 100\,G in some active areas at $z = 0$. 
Long field lines linking the two active regions definitely exist, when starting from both altitudes. At the photospheric level, we notice that the connectivities have various sizes and magnitudes, the smallest (and faintest) ones corresponding to local dipoles in the quiet Sun. On the contrary, at $z = 10$\,Mm, these local dipoles disappear, probably because of their small size, and consequently altitude. One recognizes the essential long range connectivity between the two AR, along with an internal connectivity within AR\,10942.
 This large scale connectivity between the two AR has already been found using different models by \citet{2007Sci...318.1585S}, who did not discuss it, and by \citet{2008ApJ...676L.147H}.

\subsection{Estimation of the mass flows}
\label{sec:flows}

\begin{figure*}[htbp]
  (a)
  \includegraphics[width=\linewidth]{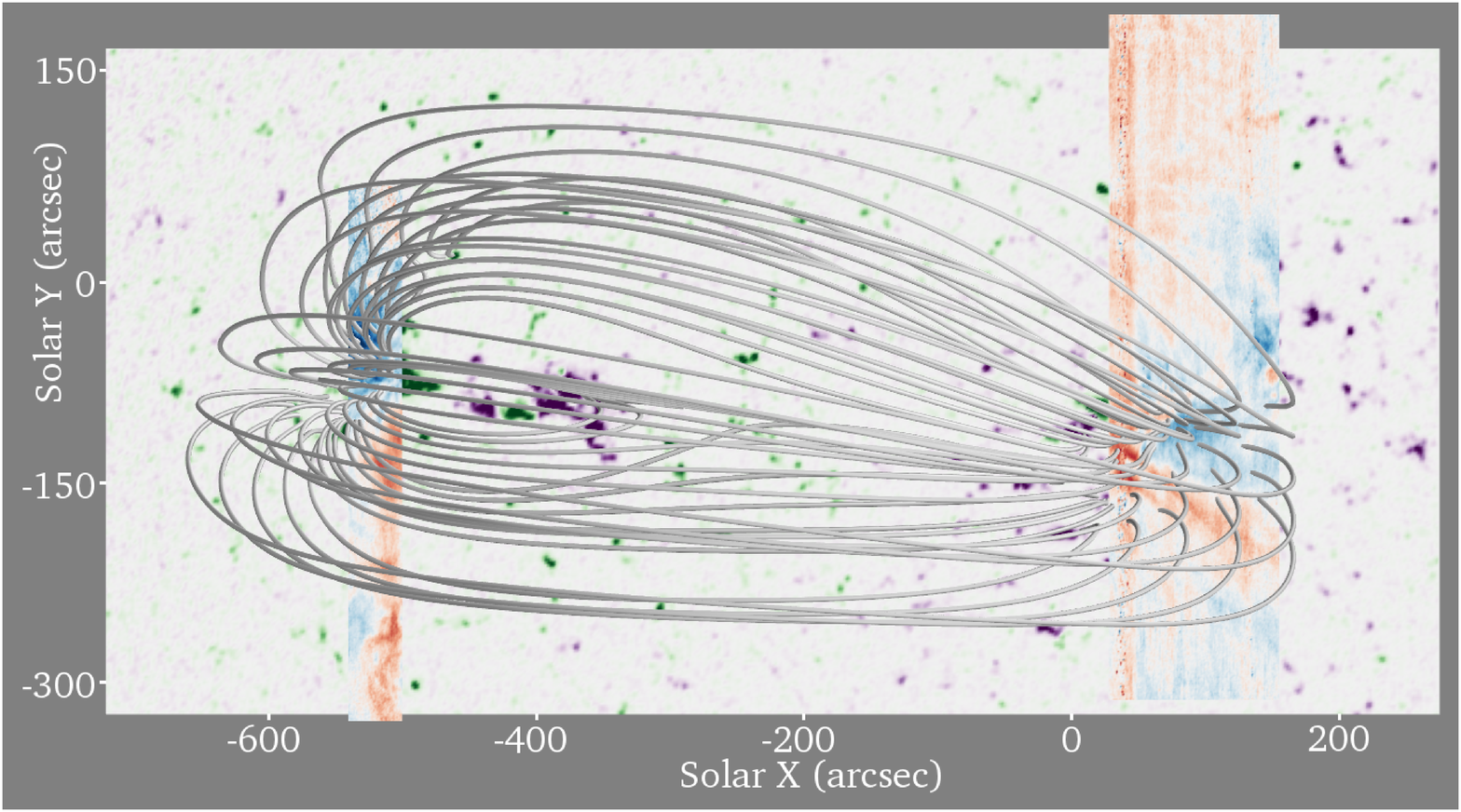}\\[.02\linewidth]
  (b)
  \includegraphics[width=.45\linewidth]{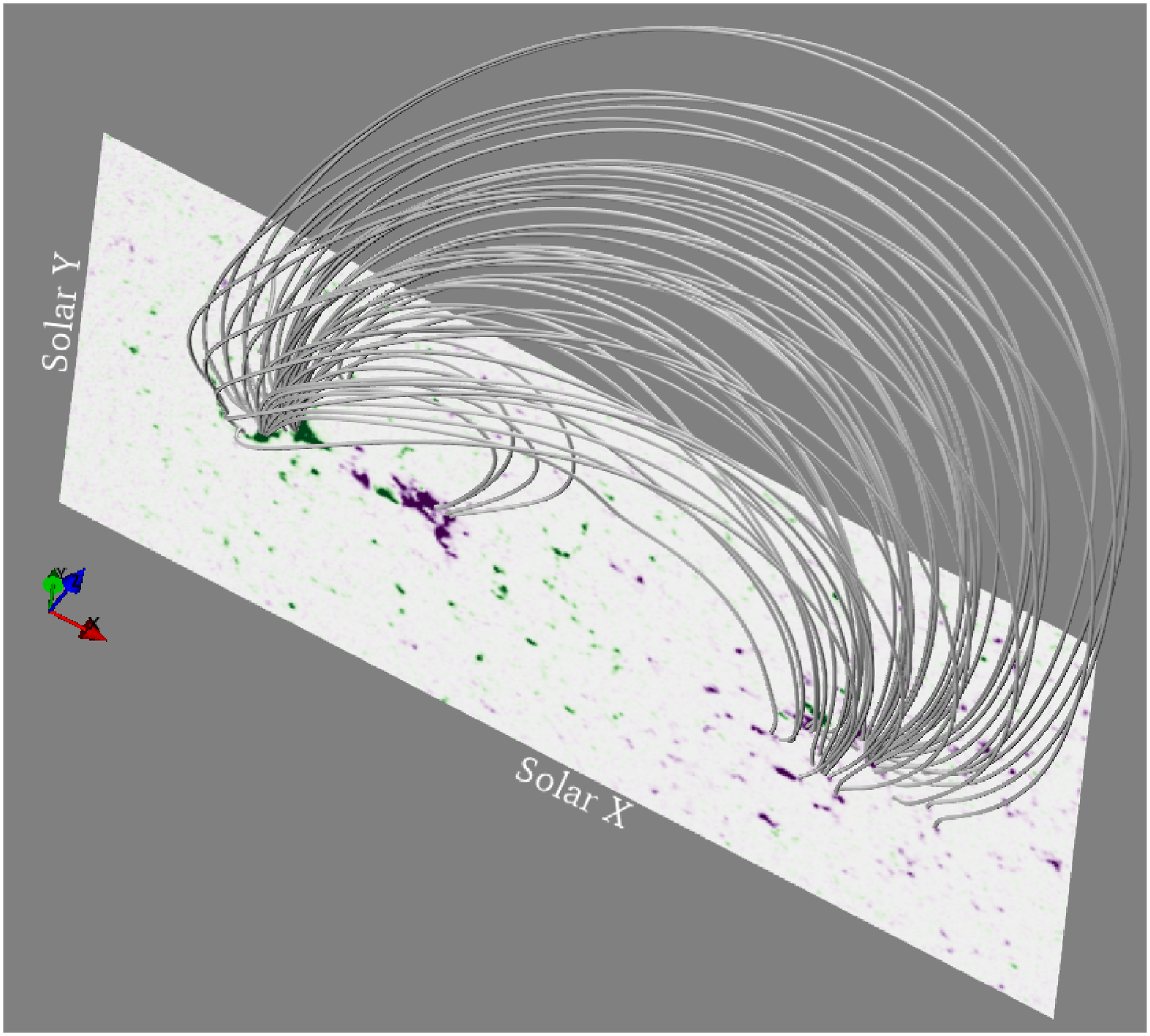}\hfill
  (c)\hspace{-10 mm}
  \includegraphics[width=.50\linewidth,bb={0 0 510 297}]{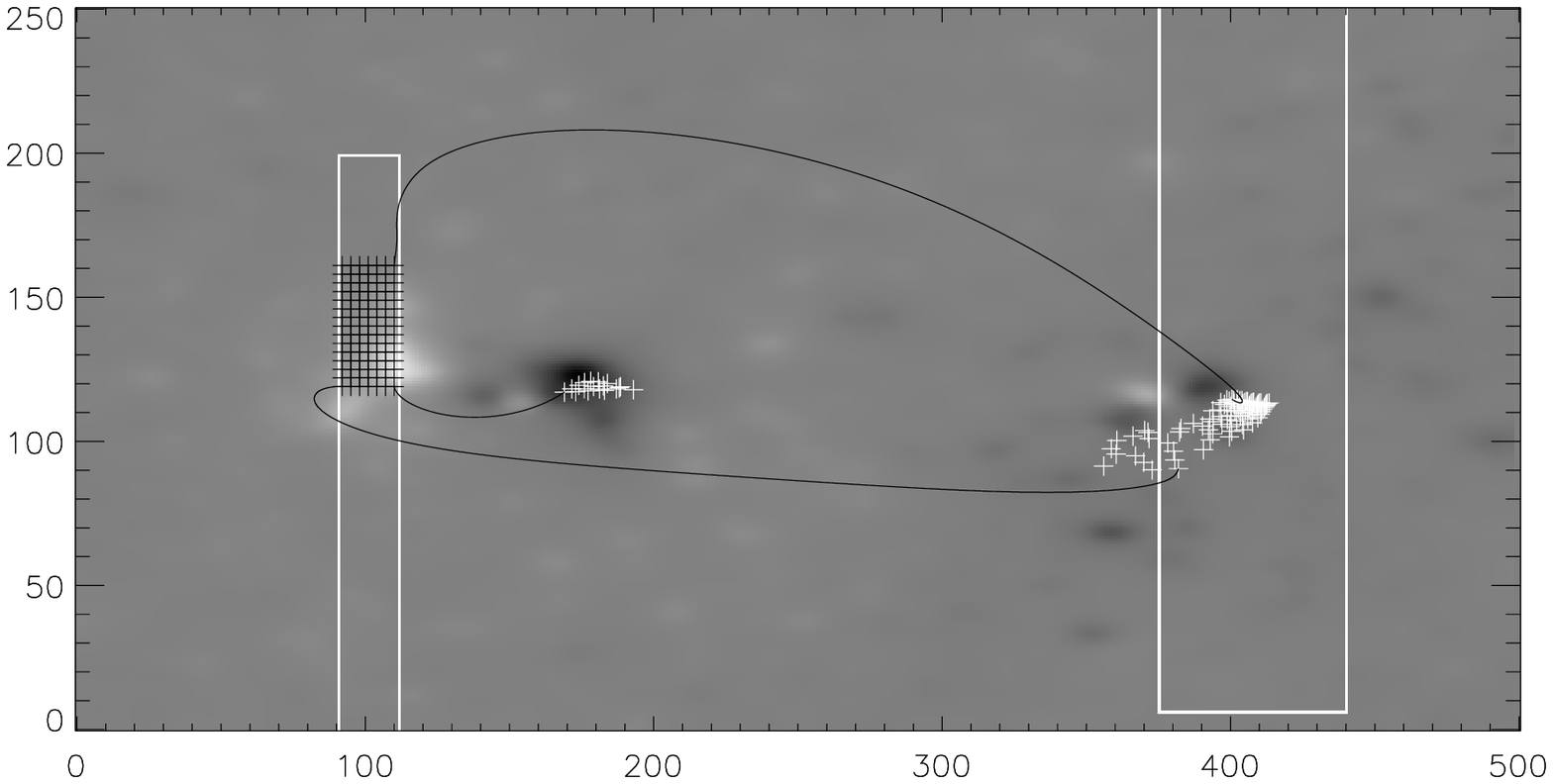}
  \caption{\label{fig:extrapol}Some selected magnetic field lines from a potential magnetic field extrapolation, with the MDI image as background: (a) Viewed from above with superimposition of the velocity maps from EIS for AR\,10942 and AR\,10943; (b) Viewed from a point of view in the Southern hemisphere; (c) Start (black crosses) and end (white crosses) points of the field lines used in Sec.~\ref{sec:flows} on a map of $B_z$ at the 10\,Mm altitude  obtained from the extrapolation; three selected lines are drawn in black. The black frames are the EIS FOV.}
\end{figure*}

We explore further the magnetic field connection between the ARs, using the potential field extrapolation (Fig.~\ref{fig:extrapol}a and b) and EIS spectral scans. The co-alignement between maps of EIS intensity and maps of the extrapolated magnetic field is made by using cross-correlation between MDI (8:05\,UT) and STEREO \ion{Fe}{12} 195 (8:05\,UT) image and cross-correlation between STEREO \ion{Fe}{12} 195 image and EIS \ion{Fe}{12} 195.12 (respectively at 05:47\,UT for AR\,10943 and at 11:16\,UT for AR\,10942).

For each AR, we have measurements of the line-of-sight velocity $v_s$ (oriented downwards), and the signed mass flux through a surface $S$ orthogonal to the line-of-sight is
\begin{equation}
  F=\iint_S \rho \vec{v}\cdot d^2 \vec{S}
\end{equation}
which reduces to $\rho v_s S$ if $S$ is small enough.
Note that $F$ has the same sign as $v_s$.

\begin{figure}[htbp]
  \includegraphics[width=1.\linewidth]{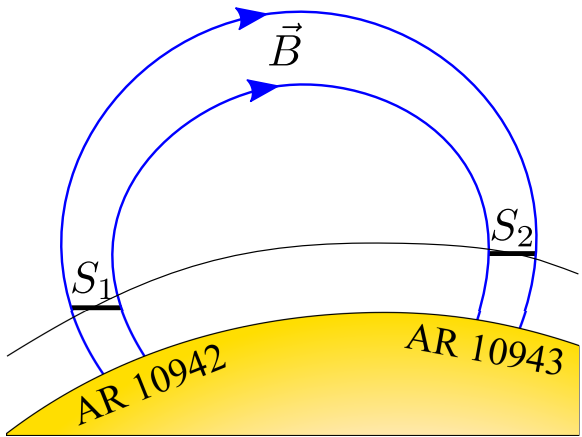}

  \caption{ \label{fig:flux} Cartoon of the magnetic flux tube connecting AR\,10942 and AR\,10943. $S_1$ and $S_2$ are the cross-sections perpendicular to the line-of-sight of the flux tube at the observed altitude (we chose 10\;Mm for \ion{Fe}{12}), above AR\,10942 and AR\,10943 respectively. The magnetic field in cross-section $S_i$ is $\vec B_i$, the density is $\rho_i$, and the velocity is $\vec v_i$. }
\end{figure}

We choose to consider the two sections $S_1$ and $S_2$ of the flux tube represented in Fig.~\ref{fig:flux}. We decompose $S_1$ into small areas $S_{1i}$  and then we define $S_{2i}$ by the intersection with $S_2$ of the flux tube originating from $S_{1i}$: we have $S_1 = \sum_i S_{1i}$ and $S_2 = \sum_i S_{2i}$.

The ratio of the total mass flux through $S_2$ to the total mass flux through $S_1$ is
\begin{align}
  \label{eq:ratio}
  R
  = \frac{F_2}{F_1}
  = \frac{\sum_i F_{2i}}{\sum_i F_{1i}}
  = \frac{\sum_i \rho_{2i} v_{s2i} S_{2i}}{\sum_i \rho_{1i} v_{s1i} S_{1i}}
\end{align}

We use the conservation of the magnetic flux $\phi_i$ in each of these small flux tubes\footnote{Flux and magnetic field are in absolute values because the surfaces' orientation is given by the line-of-sight instead of the direction of $\vec B$.}:
\begin{equation}
  |\phi_i| = |B_{s1i}|S_{1i} = |B_{s2i}| S_{2i}
\end{equation}
and we decompose $S_{2i}$ into $S_{1i} S_{2i} / S_{1i}$, which gives $S_{2i} = S_{1i} |B_{s1i} / B_{s2i}|$. Finally, as all $S_{1i}$ are chosen with the same area (a MDI pixel), we obtain the mass flux ratio
\begin{equation}
  R = \frac{\sum_i \rho_{2i} v_{s2i} |B_{s1i} / B_{s2i}|}{\sum_i \rho_{1i} v_{s1i}}
\end{equation}

In this equation, the density and the Doppler velocity from the EIS \ion{Fe}{12} spectra in AR\,10942 and 10943 (see Sec.~\ref{sec:secobs}) are derived with a $3\times3$-pixel median filter. We consider that the \ion{Fe}{12} lines are emitted at an altitude of 10\,Mm in hot loops. The values of the line-of-sight magnetic fields $B_{1i}$ and $B_{2i}$ are taken from the footpoints of magnetic field lines starting at an altitude of 10\,Mm on a $7\times15$-point grid in the outflow at the edge of AR\,10942 (Fig.~\ref{fig:extrapol}c), and computed until they reach again the same altitude of 10\,Mm. 76 among 105 of these lines link AR\,10942 and AR\,10943 and fall inside the EIS FOV on AR\,10943.

We consider these lines to be representative of the flux tubes connecting AR\,10942 to AR\,10943. The modulus of the mass flux ratio (AR\,10943 over 10942, see Eq.~\ref{eq:ratio}) is then 0.18.

We compute the uncertainty on the mass flux ratio induced by the uncertainties on the observations by
\begin{equation}
 \frac{\delta R}{R}= \frac{\delta F_1}{|F_1|}+ \frac{\delta F_2}{|F_2|}
\end{equation}
with
\begin{align}
 \frac{\delta F_1}{|F_1|}&
 =\frac{\frac1{\sqrt{n}} \sum_i |F_{1i}| \frac{\delta F_{1i}}{|F_{1i}|}}{|F_1|} \notag \\
 &=\frac1{|F_1|\sqrt{n}}\sum_i |v_{s1i}| \rho_{1i} \left(\frac{\delta \rho_{1i}}{\rho_{1i}}+\frac{\delta v_{s1i}}{
 |v_{s1i}|}\right)
\end{align}
and 
\begin{align}
 \frac{\delta F_2}{|F_2|}&
 =\frac{\frac1{\sqrt{n}} \sum_i |F_{2i}| \frac{\delta F_{2i}}{|F_{2i}|}}{|F_2|} \notag \\
 &=\frac1{|F_2| \sqrt{n}} \sum_i
    \rho_{2i} |v_{s2i} B_{s1i} / B_{s2i}| \times \\
 &   \qquad\left(
      \frac{\delta\rho_{2i}}{\rho_{2i}}+\frac{\delta v_{s2i}}{|v_{s2i}|} + \frac{\delta B_{s1i}}{|B_{s1i}|} + \frac{\delta B_{s2i}}{|B_{s2i}|}\right) . \notag
\end{align}

We have chosen a uniform
$\delta v$ of $2$\,km.s$^{-1}$ as we have an uncertainty of about $\pm 1$\,km.s$^{-1}$ for the Doppler velocity from each line fit,
 $\delta \rho=3.4\times 10^8$\,cm$^{-3}$ from the uncertainties on the intensities of the lines used to deduce the density (Fig.\ref{fig:density_ratio}),
and $\delta B=10$\,G, a value below which coherent small-scale patterns disappear in the quiet Sun. The relative uncertainty $\delta R / R$ on the mass flux ratio is then found to be 0.30.

To evaluate the uncertainty due to the co-alignment between MDI data and EIS spectra, we compute the mass flux ratio with a co-alignment shifted by one MDI pixel (1.98\,arcsec) in the two X and Y directions with respect to EIS data. The mass ratios obtained vary by a factor 2. We can then state that the co-alignment is the main cause of uncertainty.

%% file: discussion.tex
\section{Discussion}
\label{sec:discussion}

The extrapolation clearly shows us how AR\,10942 and AR\,10943 are linked.
From the observation of both ARs, it appears that the mass flow to AR\,10943 seems to be 18\% of the upflows from AR\,10942.
We can therefore conclude that the flow leaving the Sun among the upflows from AR\,10942 is at most 82\% of the measured upflows. Some of these upflows could be confined somewhere else.

\begin{figure}
\centering
 \includegraphics[width=1\linewidth]{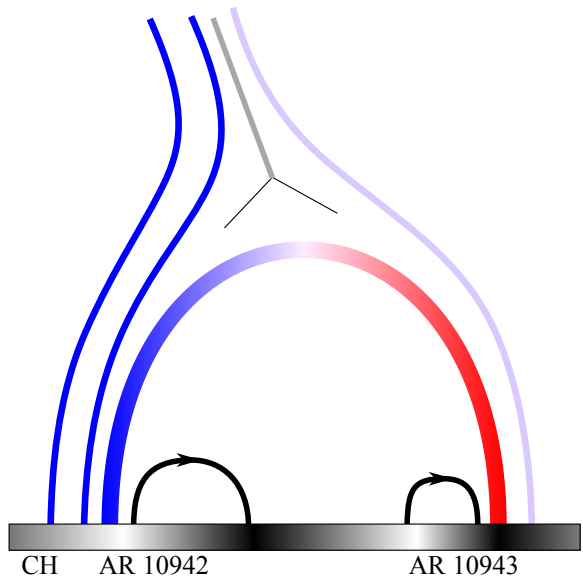}
  \caption{\label{fig:schema} Cartoon of the magnetic field and flows in the case of a Y point. The colors of the lines show the Doppler shifts that would be observed from above, with the usual convention: blue for upflows and red for downflows. The Y point is located above the colored loop and is shown  with the current sheet in grey. Some field lines are indicated with arrows.
At the bottom, the magnetic polarities coded in the usual white and black convention correspond to the location of the two ARs and the eastern coronal hole.}
\end{figure}

Another solution is not explored by our extrapolation: the presence of a Y point between the two ARs.
An argument in favour of this hypothesis is the difference between the velocities at the AR\,10942 edge and the AR\,10943 edge. Indeed, in the model proposed by \citet{2011A&A...526A.137D}, the velocity is higher for the strongest magnetic bipoles than in the faintest pole (a Coronal Hole in their model).
In this case, the flows are not directly linked and they interact with each other through an X point. In our case we clearly have open field lines, with the presence of two coronal holes, one located at the Eastern part of AR\,10942 and --- with less evidence --- the other one in the Western part of AR\,10943, which then would lead to a  high-altitude Y point (Fig.~\ref{fig:schema}). As a consequence, the measured upflows in the Eastern part include flows of material which escapes towards the Y point and the corona and cannot be found at the foot of the connecting loop in AR\,10943.
It is clear that the potential extrapolation does not allow to explore this possibility.

\section{Conclusion}
\label{sec:conclusion}
In summary, we observed Doppler-shifted structures at the edges of two Active Regions which we find to be connected with each other by large scale loops, a result we discuss in more details below.
With the advent of the EUV spectrometer EIS on Hinode, special attention has been paid on high temperature flows around coronal holes and active regions. AR\,10942 has been extensively studied (e.g. \citealt{2009ApJ...705..926B}) and it has been suggested that the strong observed upflow at its boundary could be one of the sources of the solar wind. In order to properly assess this critical result, we have investigated the possibility that the supposedly open field in this area could be actually (at least partially) connected to the preceding AR\,10943 with the consequence that part of the mass outflow from AR\,10942 could turn into downflow in AR\,10943.
In order to investigate this possibility, we analyzed the EIS spectra obtained in AR\,10943 with special attention to the \ion{Fe}{12} (195\,\AA) line formed at $\log T = 6.1$.

The EIS spectra are difficult to analyze for two reasons: most lines are blended and their profiles are affected by a strong (presumably thermal-induced) deformation during any orbit. The blends are properly dealt with the use of multi-gaussian fits. As far as the second (orbital) correction is concerned, the EIS software allows to take into account the relative shift of the profile over the orbit but leaves open the issue of the absolute reference. 
In order to overcome this difficulty, we have designed and used a technique which relies upon the cool and optically thick 256\,\AA\ \ion{He}{2} line, consequently less sensitive to activity and induced flows. The technique is described in Appendix~\ref{sec:app}.

The analysis of the velocity field in AR\,10943, performed with this technique, shows a significant feature located in the southwestern part where the downflow velocity reaches 16\,km.s$^{-1}$. It should be pointed out again that this pattern is confirmed in other lines, as mentioned in Sect.~\ref{sec:secobs}, with the noticeable exception of the \ion{Fe}{13} line at 196.5\,\AA\ which is a very weak line \citep{2009A&A...495..587Y}. At very high temperature (\ion{Ca}{17} 192.82\,\AA), it seems that the pattern is the same except for the core of the active region which is blueshifted. Moreover, through the use of the two \ion{Fe}{12} lines (196.6\,\AA\ and 195.1\,\AA) and their intensity ratio, we were able to build a density map where the downflow region density reaches a value of $5.10^{8}$ cm$^{-3}$.

We then proceeded to establish the connectivity between the two regions, and especially between the blueshifted area of AR\,10942 and the redshifted area of AR\,10943. A potential extrapolation, rather suitable for our stable active regions, was used. The connectivity between the two areas was well established although the computed field lines do not clearly appear as inter-connecting loops in the \ion{Fe}{12} image (Fig.~\ref{fig:stereo}), presumably a result of their low densities.
With the values of the local density and line-of-sight velocity in the two areas, we could compute and compare the mass flows between the respective feet of the magnetic field lines. We concluded that only 18\% of the outgoing flow from AR\,10942 is retrieved as downward flow in AR\,10943, as computed along the extrapolated field lines.
We elaborated upon the possibility of other paths followed by the upflows of AR\,10942, for various reasons: one of them being the presence of open fields in AR\,10942, which our extrapolation does not take into account; another one being the possibility of a breach of connectivity due, e.g. to a Y point where reconnection occurs \citep{2011A&A...526A.137D}.

As a conclusion, we can state that some material is exchanged between the two active regions, especially from the area of the upflow but that 82\% of the upflow from AR\,10942 is missing upon arrival on AR\,10943. Actually, these flows could be a condition for the thermal sustainability of such long loops during such a long time span. However, we cannot go further because of the many factors and uncertainties involved in the observations, whether the wavelength (velocity) reference, the non-simultaneity of the two AR observations with EIS, the reduced fields of view, or even the validity of the potential field extrapolation.

We strongly recommend that further similar studies rely upon force-free magnetic field extrapolation made possible by simultaneous SOT/Hinode observations and also closer-in-time observing modes focused on the active regions involved. We also suggest to use the reference wavelength method described in the Annex.

\acknowledgments
The authors thank the referee for her/his useful comments and suggestions. They also thank Helen Mason, Giulio Del Zanna, Durgesh Tripathi and Leon Golub for advice on spectroscopic and imaging issues.
CHIANTI is a collaborative project involving researchers at NRL (USA) RAL (UK), and the Universities of: Cambridge (UK), George Mason (USA), and Florence (Italy). 
Hinode is a Japanese mission developed and launched by ISAS/JAXA, with NAOJ as domestic partner and NASA and STFC (UK) as international partners. It is operated by these agencies in co-operation with ESA and NSC (Norway).
SOHO is a project of international cooperation between ESA and NASA.
The SECCHI data used here were produced by an international consortium of the Naval Research Laboratory (USA), Lockheed Martin Solar and Astrophysics Lab (USA), NASA Goddard Space Flight Center (USA), Rutherford Appleton Laboratory (UK), University of Birmingham (UK), Max-Planck-Institut for Solar System Research (Germany), Centre Spatiale de Li\`ege (Belgium), Institut d'Optique Th\'eorique et Appliqu\'ee (France), and Institut d'Astrophysique Spatiale (France).
The authors acknowledge the CNES and the CNRS/INSU/Programme Soleil-Terre for financial support.

\appendix
\section{Wavelength reference for EIS}
\label{sec:app}

\begin{figure}
\includegraphics[width=1\linewidth]{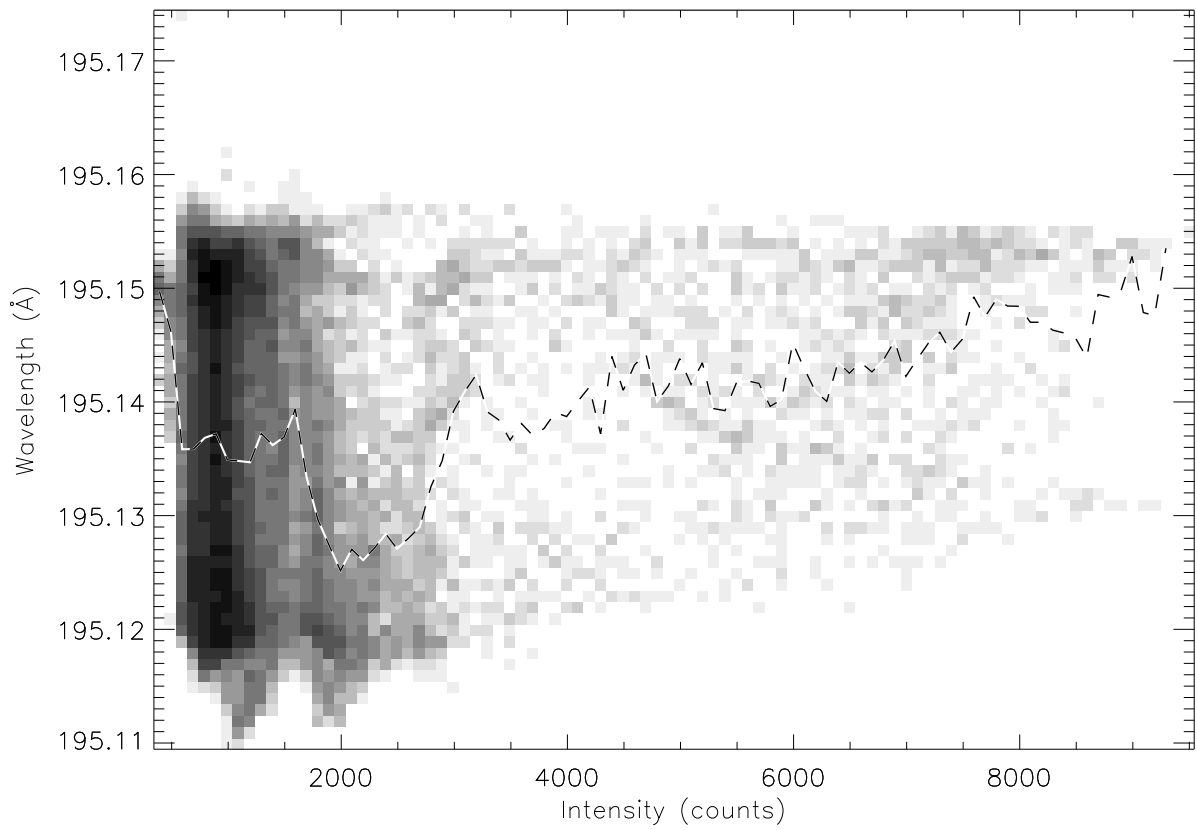}

  \caption{\label{fig:disp_fe_ncorr} Dispersion of the centroid wavelength of the \ion{Fe}{12} line profile depending on the intensity (counts) before the orbital correction.}
\end{figure}

The temperature variation along the orbit of Hinode deforms the optical axis of EIS in the dispersion plane \citep{2007PASJ...59S.865B}. The determination of the Doppler velocity is therefore more difficult because of the lack of absolute reference. Figure \ref{fig:disp_fe_ncorr} shows the wide range (0.05 \,\AA) of the \ion{Fe}{12} wavelengths given by EIS before any correction. Therefore a correction is necessary.
It has been observed \citep{2007PASJ...59S.865B} that this orbital variation of the wavelength is not the same from one orbit to the other and depends on several parameters (Hinode pointing, Earth's seasons, etc.). Thus, it is not predictable and the orbital correction, based on the considered science data set, has to be
empirical, although a correction where housekeeping data are used \citep{2010SoPh..266..209K}
is possible.
 Nevertheless, as the orbital variation is the same for both detectors, we can use the observation of any line to correct the others \citep{2007PASJ...59S.713M}.

The measured wavelength depends on the physical parameters on the Sun and the instrumental bias including the orbital variation. We want to separate these two components of the wavelength line centroid. We assume that the Doppler shifts are globally null on average in the Y dimension while the orbital variation is the same on each pixel along the Y dimension of the slit at a given time. So we can isolate the orbital variation by integrating on the Y axis. In order to smooth any residual Doppler shifts, we use a spline depending on time.

The procedure consists in averaging the line profiles along the Y axis in the quiet Sun pixels and then to fit these integrated lines with a Gaussian in order to obtain the mean wavelength of the line centroid depending on the time $\lambda_{av}(t)$.
The line centroids $\lambda_{av}(t)$ obtained are fitted with a spline $\mathcal{S}(t)$)  in order to keep only the orbital variation. The correction, that is to say the difference between the spline and the reference wavelength $\lambda_0$ (as given in CHIANTI), is denoted $\delta\lambda(t)$.

The usual way of finding the wavelength reference consists in following the above procedure. However, the issue of the line shift in the corona is not solved today. From SUMER/SOHO, it was derived that the redshift observed in the transition region (with a maximum of 30 km.s$^{-1}$ at logT= 5.3) decreased at high coronal temperatures but remained a redshift (e.g. \citealt{1998ApJS..114..151C}). Then also with SUMER data, \citet{1999ApJ...522.1148P} showed that at high temperatures, one actually had a blueshift, i.e. upward velocities. Consequently, we suspect that taking a hot line as a wavelength reference, even with a substantial spatial averaging, may not be a safe procedure.

As the \ion{He}{2} 256.32\,\AA\ line is the coldest line in the EIS wavelength ranges (being formed from the upper chromosphere to the lower transition region) and is optically thick, the line is less affected by the solar flows than the other EIS lines. Therefore, the centroid of its profile is an efficient reference. The drawback is that \ion{He}{2} 256.32\,\AA\ is blended with hot lines.
In the quiet Sun, \ion{He}{2} 256.32\,\AA\ is dominant (its contribution is more than 80\% of the blend \citep{2007PASJ...59S.857Y} and can be separated from the blending lines with a two-Gaussian fit. For the active Sun, it is more difficult to separate \ion{He}{2} 256.32\,\AA\ from its blends, as they are hot lines and their contribution is changing depending on the observed structures.

In our study, for example, we choose to apply the correction obtained from \ion{He}{2} 256.32\,\AA\ on the \ion{Fe}{12} 196.64\,\AA\ line. We computed $\delta\lambda_{HeII}(t)$, we fitted $\lambda_{FeXII}(y,t)$ by a gaussian fit and we corrected each pixel following the formula $\lambda_{FeXII corrected}(y,t)=\lambda_{FeXII}(y,t)-\delta\lambda_{HeII}(t)$ 

As \ion{He}{2} 256.32\,\AA\ is essentially blended with \ion{Si}{10} 256.37\,\AA\ \citep{2005AdSpR..36.1503D}. We consider that we can fit these two lines only.
Because \ion{Si}{10} 261.04\,\AA\ and \ion{Si}{10} 256.37\,\AA\ have the same lower level ($2s 2p^{2\: 2}P_{1/2}$), we can use the branching ratio between \ion{Si}{10} 261.04\,\AA\ and \ion{Si}{10} 256.37\,\AA\ in order to evaluate the contribution of \ion{Si}{10} 256.37\,\AA\ in the 256\,\AA\ blend. The branching ratio denoted $\alpha$ is determined by the ratio of the intensities given by CHIANTI.

The procedure is as follows :
\begin{itemize}
\item At first, we fit \ion{Si}{10} 261.04\,\AA\ with a simple Gaussian. We obtain $\lambda _{Si\,X 261}$, $I_{Si\,X 261}$,$FWHM_{Si\,X 261}$ and the background.
\item We transpose the characteristics we obtained for \ion{Si}{10} 261.04\,\AA\ to \ion{Si}{10} 256.37\,\AA, that is to say for the wavelength,   $\lambda _{Si\,X 256}= \lambda _{Si\,X 261} - \lambda _{0,Si\,X 261} + \lambda _{0,Si\,X 256}$, we consider the same width, and for the intensity,  $I_{256}=\alpha I_{261}$ where $\alpha$ is the branching ratio.
\item We fit \ion{He}{2} 256.32\,\AA\ and its blend with a 5-parameter fit ($\lambda_{He\,II}$, $I_{He\,II}$, $I_{Si\,X 256}$, $FWHM_{He\,II}$ and the intensity background), with fixed value for $\lambda _{Si\,X 256}$ and $FWHM_{Si\,X 256}$.
\item We extract the correction $\delta\lambda(t)$ from $\lambda_{He\,II}(t)$ as described previously and subtract it from the wavelength obtained for the studied line.  
 \end{itemize}

\begin{figure}
\includegraphics[width=1\linewidth]{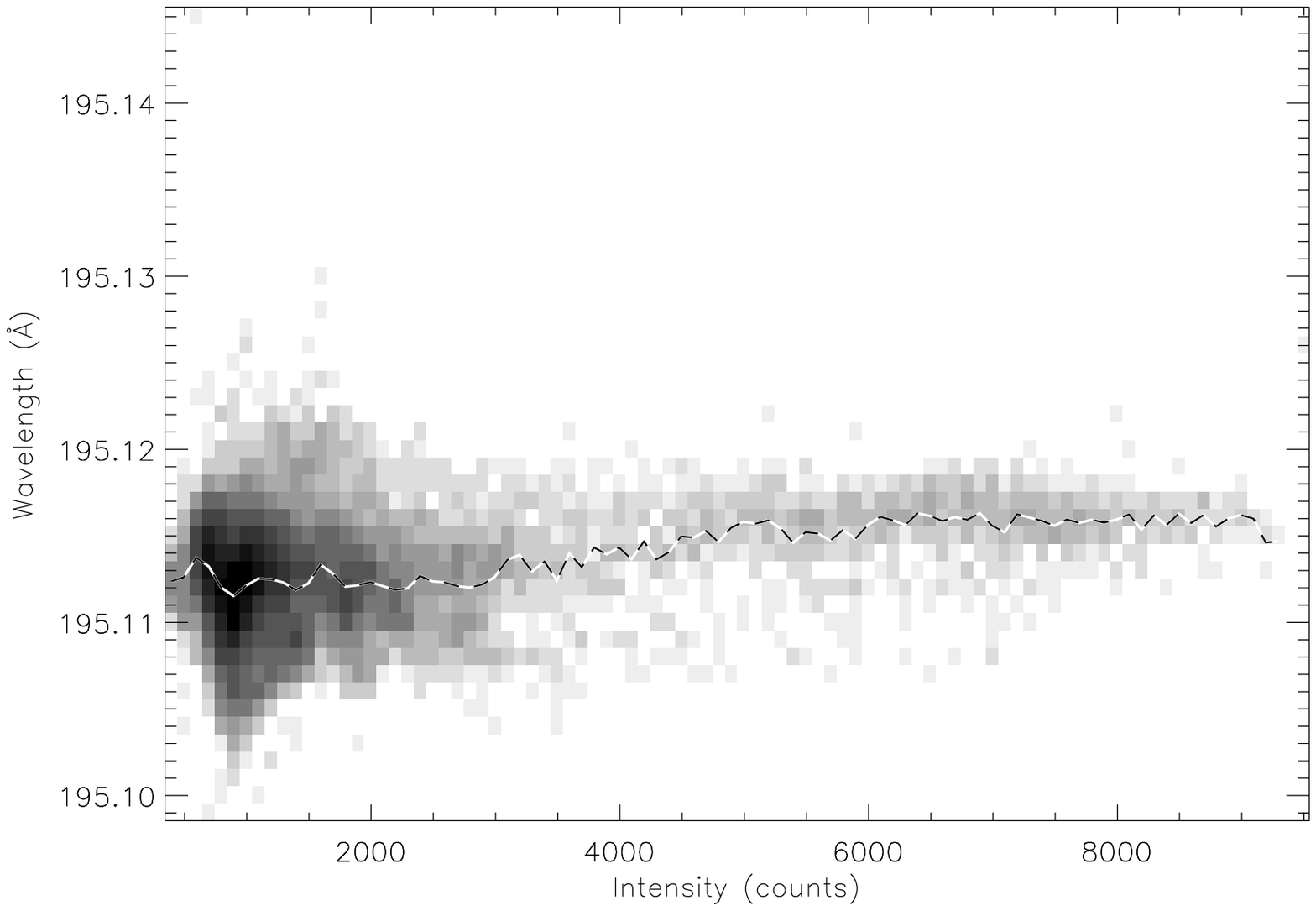}

  \caption{\label{fig:disp_fe_corr256} Dispersion of the centroid wavelength of the \ion{Fe}{12} line profile depending on the intensity (counts) after the orbital correction using \ion{He}{2}.}
\end{figure}

\begin{figure}
\includegraphics[width=1\linewidth]{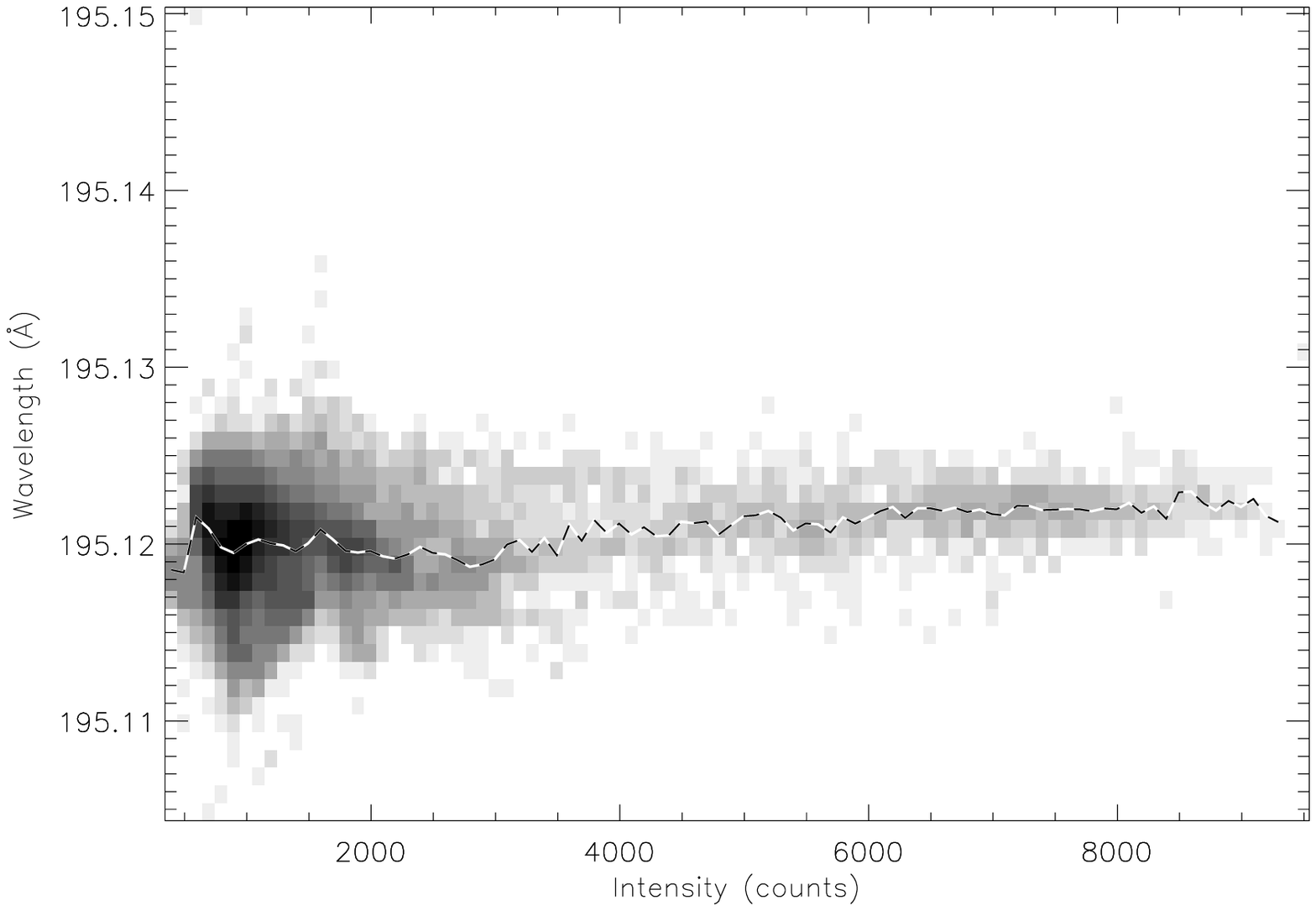}

  \caption{\label{fig:disp_fe_corr195} Dispersion of the centroid wavelength of the \ion{Fe}{12} line profile depending on the intensity (counts) after the orbital correction using \ion{Fe}{12}.}
\end{figure}

Figure \ref{fig:disp_fe_corr256} shows the dispersion of the wavelength of \ion{Fe}{12} 195.12\,\AA\ after the correction proposed in this appendix. We can note that the wavelength increases slightly with the intensity, a result in agreement with the fact that the Active Region is slightly redshifted. The orbital correction computed with the SolarSoft library based on the \ion{Fe}{12} 195.12\,\AA\ line shows the same tendency (Fig.\ref{fig:disp_fe_corr195}).
There is a systematic difference of 0.008\,\AA\ between the two corrections which is consistent with the line shift depending on the temperature discussed in \citet{1998ApJS..114..151C} and in \citet{1999ApJ...522.1148P}. They reported a redshift of 5.3 km.s$^{-1}$ for \ion{Si}{3} $\lambda$ 1206.51+$\lambda$ 1206.53\,\AA\  \citep{1998ApJS..114..151C} at about the same temperature as \ion{He}{2} 256 ($\log T_{max}=4.7$) and a blueshift of 6 km.s$^{-1}$ for \ion{Fe}{12} 1349\,\AA\ line \citep{1977ApJ...214..898S} which corresponds to a total relative Doppler shift of 0.008\,\AA. Therefore, we apply a redshift of 0.008\,\AA\ on our Doppler shifts analysis (Sec. \ref{sec:secobs} and Sec.~\ref{sec:flows}), which implies that we adopt an average quiet Sun \ion{Fe}{12} line line at rest.

